\documentclass[nohyper,12pt,letterpaper]{JHEP3}
\usepackage{epsfig}
\usepackage[latin1]{inputenc}
\usepackage{bbm}

\author{A. Mauri$^{1}$, S. Penati$^{2}$, M. Pirrone$^{2}$, A. Santambrogio$^{1}$ and
D. Zanon$^{1}$\\
${}^{1}$Dipartimento di Fisica, Universit\`a di Milano and INFN, Sezione di Milano,
Via Celoria 16, I-20133 Milano, Italy\\
\qquad\\
${}^{2}$Dipartimento di Fisica, Universit\`a di Milano--Bicocca and INFN, Sezione di Milano--Bicocca,
Piazza della Scienza 3, I-20126 Milano, Italy\\
\qquad\\
E-mail: \email{andrea.mauri@mi.infn.it, silvia.penati@mib.infn.it, marco.pirrone@mib.infn.it,
alberto.santambrogio@mi.infn.it, daniela.zanon@mi.infn.it}}

\abstract{For $\mc N=1$ $SU(N)$ SYM theories obtained as marginal deformations of the $\mc N=4$
parent theory we study perturbatively some sectors of the chiral ring in the weak coupling
regime and for finite $N$. By exploiting the relation between the definition of chiral ring and
the effective superpotential we develop a procedure which allows us to easily determine protected
chiral operators up to $n$ loops once the superpotential has been computed up to $(n-1)$ order.
In particular, for the Lunin--Maldacena $\b$--deformed theory we determine the quantum structure
of a large class of operators up to three loops. We extend our procedure to more general
Leigh--Strassler deformations whose chiral ring is not fully understood yet and determine
the weight--two and weight--three sectors up to two loops. We use our results to infer general
properties of the chiral ring.}

\preprint{Bicocca-FT-06-8 \\
IFUM-865-FT}

\title{On the perturbative chiral ring for marginally deformed  $\mc N=4$ SYM theories}


\csname @addtoreset\endcsname{equation}{section}

\def\bseq{\begin{subequation}}  % = 1a 1b
\def\eseq{\end{subequation}}
\def\bsea{\begin{subeqnarray}}  % = 1.1a 1.1b
\def\esea{\end{subeqnarray}}
\def\Bar#1{\overline{#1}}                       % big bar

\newcommand{\beq}{\begin{equation}}
\newcommand{\bea}{\begin{eqnarray}}
\newcommand{\eea}{\end{eqnarray}}
\newcommand{\eeq}{\end{equation}}

\newcommand {\non}{\nonumber}

\newcommand{\mc}{\mathcal}

\renewcommand{\a}{\alpha}
\renewcommand{\b}{\beta}

\renewcommand{\d}{\delta}

\newcommand{\g}{\gamma}

\newcommand{\D}{\Delta}
\newcommand{\e}{\epsilon}

\renewcommand{\l}{\lambda}

\renewcommand{\o}{\omega}

\newcommand{\Db}{\overline{D}}

\newcommand{\Phib}{\overline{\Phi}}

\newcommand{\qb}{\bar{q}}

\def\Mb{\kern 2pt\mathchoice
        {%displaystyle
         \vbox{\hrule width10pt height 0.4pt depth 0pt
         \kern 1.2pt\hbox{\kern -2pt$\displaystyle M$}}}
        {%textstyle
         \vbox{\hrule width10pt height 0.4pt depth 0pt
         \kern 1.2pt\hbox{\kern -2pt$\textstyle M$}}}
        {%scriptstyle \kern 0.5pt
\vbox{\hrule width6pt height 0.4pt depth 0pt
         \kern 1.0pt\hbox{\kern -2pt$\scriptstyle M$}}}
        {%scriptscriptstyle \kern 0.5pt
         \vbox{\hrule width5pt height 0.4pt depth 0pt
         \kern 0.8pt\hbox{\kern -2pt$\scriptscriptstyle M$}}}}

\def\Sb{\kern 2pt\mathchoice
        {%displaystyle
         \vbox{\hrule width6pt height 0.4pt depth 0pt
         \kern 1.2pt\hbox{\kern -2pt$\displaystyle S$}}}
        {%textstyle
         \vbox{\hrule width6pt height 0.4pt depth 0pt
         \kern 1.2pt\hbox{\kern -2pt$\textstyle S$}}}
        {%scriptstyle
         \vbox{\hrule width3.5pt height 0.4pt depth 0pt
         \kern 1.0pt\hbox{\kern -2pt$\scriptstyle S$}}}
        {%scriptscriptstyle
         \vbox{\hrule width3pt height 0.4pt depth 0pt
         \kern 0.8pt\hbox{\kern -2pt$\scriptscriptstyle S$}}}}

\def\Rb{\kern 2pt\mathchoice
        {%displaystyle
         \vbox{\hrule width5.5pt height 0.4pt depth 0pt
         \kern 1.2pt\hbox{\kern -2.5pt$\displaystyle R$}}}
        {%textstyle
         \vbox{\hrule width5.5pt height 0.4pt depth 0pt
         \kern 1.2pt\hbox{\kern -2.5pt$\textstyle R$}}}
        {%scriptstyle
         \vbox{\hrule width3.5pt height 0.4pt depth 0pt
         \kern 1.0pt\hbox{\kern -2.2pt$\scriptstyle R$}}}
        {%scriptscriptstyle
         \vbox{\hrule width3pt height 0.4pt depth 0pt
         \kern 0.8pt\hbox{\kern -2.2pt$\scriptscriptstyle R$}}}}

  \def\pp{{\mathchoice
      {%displaystyle
          \kern 1pt%
          \raise 1pt
          \vbox{\hrule width5pt height0.4pt depth0pt
            \kern -2pt
            \hbox{\kern 2.3pt
              \vrule width0.4pt height6pt depth0pt
              }
            \kern -2pt
            \hrule width5pt height0.4pt depth0pt}%
            \kern 1pt
       }
        {%textstyle
          \kern 1pt%
          \raise 1pt
          \vbox{\hrule width4.3pt height0.4pt depth0pt
            \kern -1.8pt
            \hbox{\kern 1.95pt
              \vrule width0.4pt height5.4pt depth0pt
              }
            \kern -1.8pt
            \hrule width4.3pt height0.4pt depth0pt}%
            \kern 1pt
        }
        {%scriptstyle
          \kern 0.5pt%
          \raise 1pt
          \vbox{\hrule width4.0pt height0.3pt depth0pt
            \kern -1.9pt  %[e]=0.15pt
            \hbox{\kern 1.85pt
              \vrule width0.3pt height5.7pt depth0pt
              }
            \kern -1.9pt
            \hrule width4.0pt height0.3pt depth0pt}%
            \kern 0.5pt
        }
        {%scriptscriptstyle
          \kern 0.5pt%
          \raise 1pt
          \vbox{\hrule width3.6pt height0.3pt depth0pt
            \kern -1.5pt
            \hbox{\kern 1.65pt
              \vrule width0.3pt height4.5pt depth0pt
              }
            \kern -1.5pt
            \hrule width3.6pt height0.3pt depth0pt}%
            \kern 0.5pt%}
        }
    }}

  \def\mm{{\mathchoice
               {%displaystyle
                 \kern 1pt
           \raise 1pt    \vbox{\hrule width5pt height0.4pt depth0pt
                  \kern 2pt
                  \hrule width5pt height0.4pt depth0pt}
                 \kern 1pt}
               {%textstyle
                \kern 1pt
           \raise 1pt \vbox{\hrule width4.3pt height0.4pt depth0pt
                  \kern 1.8pt
                  \hrule width4.3pt height0.4pt depth0pt}
                 \kern 1pt}
               {%scriptstyle
                \kern 0.5pt
           \raise 1pt
                \vbox{\hrule width4.0pt height0.3pt depth0pt
                  \kern 1.9pt
                  \hrule width4.0pt height0.3pt depth0pt}
                \kern 1pt}
               {%scriptscriptstyle
               \kern 0.5pt
         \raise 1pt  \vbox{\hrule width3.6pt height0.3pt depth0pt
                  \kern 1.5pt
                  \hrule width3.6pt height0.3pt depth0pt}
               \kern 0.5pt}
               }}

\def\pd{{\kern0.5pt
           + \kern-5.05pt \raise5.8pt\hbox{$\textstyle.$}\kern
0.5pt}}

\def\pmd{{\kern0.5pt
          \pm \kern-5.05pt
\raise6.3pt\hbox{$\textstyle.$}\kern1.5pt}}

\def\md{{\mathchoice
   {%displaystyle
      {{\kern 1pt - \kern-6.2pt \raise5pt\hbox{$\textstyle.$}\kern
1pt}}}
    {%textstyle
      {{\kern 1pt - \kern-6.2pt \raise5pt\hbox{$\textstyle.$}\kern
1pt}}}
    {%scriptstyle
      {\kern0.5pt - \kern-5.05pt
\raise3.4pt\hbox{$\textstyle.$}\kern0.5pt}}
    {%scriptscriptstyle
      {\kern0.5pt - \kern-5.05pt
\raise3.4pt\hbox{$\textstyle.$}\kern0.5pt}}}}

\begin{document}

\section{Introduction}

The original formulation of the AdS/CFT correspondence
\cite{M,GKP,W} involves a SYM theory with maximal supersymmetry.
First steps in the direction of studying the correspondence with a
lower number of supersymmetries were undertaken in \cite{GPPZAKY}.

If $\mc N=1$ superconformal invariance is required the field theory
can be realized by orbifold constructions \cite{KS} or by the {\em exactly} marginal
deformations of the $\mc N=4$ SYM first classified in \cite{LS}. The second class
of theories has been extensively studied in a field theory approach
\cite{LS,OASR,DHK,D} and in the context of the AdS/CFT correspondence
\cite{BL,BJL,BJL3,BJL2,NP}.

The interest in marginal deformed SYM theories has recently
received a considerable boost thanks to the work of
Lunin--Maldacena \cite{LM} where the gravity dual of the so called
$\b$--deformed theory has been proposed. It corresponds to the low
energy limit of a string theory on a deformed background ${\rm
AdS}_5 \times {\rm S}_{\b}^5$ obtained by SL(2,R) transforming the
$\tau$ modulus of a two--torus inside ${\rm S}^5$.
Alternatively, it can be obtained from the original  ${\rm AdS}_5
\times {\rm S}^5$ solution by applying a TsT transformation in $S^5$
\cite{LM,F,RVY,AAF}.

A considerable effort has been devoted so far to provide tests of
the correspondence in its marginal deformed version.  As for the
${\rm AdS}_5 \times {\rm S}^5$ original correspondence,
perturbative properties of the field theory have been investigated:
For the $SU(N)$ case the condition which constrains the couplings of the
theory in order to have $\mc N=1$ superconformal invariance has been
determined perturbatively up to three loops \cite{FG,PSZ4,RSS}. In
the large $N$ limit the exact superconformal condition has been
found in \cite{MPSZ}.  Nonrenormalization properties of operators
in the chiral ring have been established perturbatively
\cite{FG,PSZ4,RSS} and multiloop amplitudes have been computed
\cite{PSZ4,RSS,K1}. The exact anomalous dimensions for spin--2
operators of the form ${\rm Tr}(\Phi_1^J \Phi_2)$ have been
determined \cite{MPSZ} for $N$, $J$ unrelated and large\footnote{The same kind of limit
has been recently considered in \cite{HM} for studying magnons in the $\mc N=4$ SYM theory.}.
Finally, the gauge one--loop effective action has been computed \cite{KT} for a
particular background configuration. Nonperturbative instantonic
effects have been also considered \cite{K3}.

Integrability properties of the original $\mc N=4$ SYM theory (see
\cite{B} for a review and list of references) survive the
$\b$--deformation \cite{R,BC,F,BR2,BM} and Bethe ansatz techniques can be used also
in this case to compute the spectrum of anomalous dimensions of
composite operators.

On the string theory side BPS states have been investigated in
\cite{BC2} for orbifold configurations. Integrability properties
have been exploited on the two sides of the correspondence in order to
match the energies of semiclassical fast rotating strings with
one--loop anomalous dimensions of scalar operators
\cite{FRT2,BDR2,R2,CK,CO}. The spectrum of states has been also investigated
in the BMN limit \cite{KMSS,MPP}.

Non--supersymmetric generalizations of the Lunin--Maldacena
$\b$--deformation have been proposed \cite{F} and further
investigations have been carried on \cite{FRT,RVY,P,FKM,O}. Very recently,
deformations obtained by acting with TsT transformations in ${\rm AdS}_5$
have been also proposed \cite{MLS}.

Finally, the Lunin--Maldacena deformation has been applied in the
context of dipole theories with the purpose of disentangling the
KK modes (whose dynamics gets affected by the deformation) from the gauge modes
\cite{GN,P2,BDR,KKLM,G}.

In a previous paper \cite{PSZ4} we have initiated the study of the
chiral ring of the $SU(N)$ $\b$--deformed SYM theory by exploiting
perturbative techniques in $\mc N=1$ superspace \cite{GRS,superspace,
PSZ1, PSZ2, PS}. There we concentrated on the single--trace sector
of the chiral ring: For the lowest dimensional scalar operators
we proved the vanishing of their anomalous dimensions up to two
loops and the appearance of finite corrections to their
correlation functions, in contradistinction to the $\mc N=4$ case. In
particular, our two--loop results confirmed the protection
\cite{FG} of the operator ${\rm Tr}(\Phi_i \Phi_j)$, $i \neq j$
which was missing in the list of CPO's of the theory
\cite{BL,BJL,LM}.

In this paper we intend to pursue our investigation and extend it
to higher dimensional sectors of the chiral ring for scalar chiral
superfields. We work at finite $N$ and take into account mixing
among sectors with different trace structures.
Exploiting the definition of quantum chiral ring we reduce the
determination of protected operators up to order $n$ in
perturbation theory to the evaluation of the effective
superpotential up to order $(n-1)$. Precisely, from the knowledge
of the effective superpotential we determine perturbatively all
the quantum descendant operators of naive scale dimension $\D_0$,
and find the CPO's as the operators which are orthogonal order by
order to the descendants.

For the $\b$--deformed theory we investigate the spin--2
sector\footnote{We use the notation of \cite{FRT2} and call
``spin--$n$'' the sector containing operators made by products of
$n$ different flavors.} and applying our procedure to simple cases
($\D_0=4,5$) we determine the protected operators up to three
loops. In the sectors we have studied we can always define 
descendant operators which do not receive
quantum corrections. This seems to be a general property of the
spin--2 operators: Despite the nontrivial appearance of finite
perturbative corrections to the effective action, the quantum
descendant operators defined in terms of the effective
superpotential coincide with their expressions given in terms of
the  classical superpotential (up to possible mixing among them).

We then investigate the spin--3 sector where, due to the
appearance of Konishi--like anomalies, we need restrict our
analysis at two loops in order to avoid dealing with mixed
gauge/scalar operators. Up to this order the descendant operators
we consider are the classical ones. However, in this sector we
expect higher order corrections to the descendants to appear
together with a nontrivial dependence on the anomaly term.
Therefore, the non--renormalization properties of the descendant
operators that we experiment for the spin--2 sector are not 
a general feature of the theory.

We generalize our procedure to the study of protected operators
for the $\mc N=1$ superconformal theory associated to the full
Leigh--Strassler deformation. Even if the gravity dual of this
theory is not known yet, it is anyway interesting to figure out
the general structure of its chiral ring. Still at finite $N$, we
study explicitly the weight--2 and weight--3 sectors up to two
loops and perform a preliminary analysis of the general sectors at
least at lowest order in the couplings. An interesting result we
find is that, because of the discrete $Z_3$ symmetries of the theory,
the sectors corresponding to conformal weights which are
multiple of 3 have a different operator structure from the other
ones.

The plan of the paper is as follows: After an introductory
section on the $\b$--deformed superconformal theory, in Section 3
we introduce the definition of perturbative chiral ring and
discuss the general procedure we adopt to determine the CPO's of
the theory. In Section 4 we compute the perturbative effective
superpotential up to two loops as required to determine protected
operators up to three loops. These are then the subject of
Sections 5 and 6 for the spin--2 and spin--3 sectors, respectively.
In Section 7 we study the more general $\mc N=1$
superconformal theory described by the full Leigh--Strassler
superpotential. Some conclusions follow plus an Appendix on loop
integrals we used in our calculations.

\section{Generalities on the $\b$--deformed theory}

Given the $\mc N=4$ SYM theory in ${\cal N}=1$ superspace notation
we consider its deformation \cite{LS,LM}
\bea\label{baction}
S
&=&\int d^8z~ {\rm Tr}\left(e^{-gV} \Phib_i e^{gV} \Phi^i\right)+
\frac{1}{2g^2}
\int d^6z~ {\rm Tr} W^\a W_\a\nonumber\\
&&+ih  \int d^6z~ {\rm Tr}( q \, \Phi_1 \Phi_2 \Phi_3 - \qb \, \Phi_1 \Phi_3 \Phi_2 )
 - i \overline{h} \int d^6\bar{z}~ {\rm Tr} ( \qb \, \Phib_1 \Phib_3 \Phib_2 - q \,
\Phib_1 \Phib_2 \Phib_3 )  \non \\
\eea
where we have set $q \equiv e^{i\pi\b}$, $\bar{q} \equiv e^{-i\pi\b}$,
$\b$ real. The gauge coupling $g$ has been chosen to be real in order to avoid
dealing with instantonic effects, whereas $h$ is generically complex.

The superfield strength $W_\a= i\Db^2(e^{-gV}D_\a e^{gV})$ is given in terms of a real
prepotential $V$ and $\Phi_{1,2,3}$ contain the six scalars of
the original ${\cal N}=4$ SYM theory organized into the ${\bf 3}\times \bf{ \bar 3}$ of $SU(3)$
subgroup of the R--symmetry group $SU(4)$. We write $V=V^aT_a$, $\Phi_i=\Phi_i^a T_a$ where
$T_a$ are $SU(N)$ matrices in the fundamental representation\footnote{For more details on our
conventions we refer to \cite{PSZ1, PSZ2, PS, PSZ4}.}.

The $\beta$--deformation breaks ${\cal N}=4$ supersymmetry to ${\cal N}=1$ and
the original $SU(4)$ R--symmetry to $U(1)_R$. However,
two extra non--R--symmetry global $U(1)$'s survive. Applying the $a$--maximization procedure \cite{IW}
and the conditions of vanishing ABJ anomalies it turns out that $U(1)_R$ is the one which assigns
the same R--charge $\o$ to the three elementary superfields, whereas the charges with respect
to the two non--R--symmetries $U(1)_1 \times U(1)_2$ can be chosen to be
$(\Phi_1,\Phi_2, \Phi_3) \rightarrow (0,1,-1)$ and $(-1,1,0)$, respectively.

The action (\ref{baction}) possesses two extra discrete symmetries. One is the
$Z_3$ associated to cyclic permutations of $(\Phi_1,\Phi_2,\Phi_3)$ which is
a remnant of the original $SU(3) \subset SU(4)$ symmetry of the undeformed theory,
whereas the other one corresponds to exchanges
\beq
\Phi_i \leftrightarrow \Phi_j~,~~ i \neq j  \qquad {\rm and } \qquad q \to -\bar{q}
\quad ( \b \to 1 - \b)
\label{qsymmetry}
\eeq
The equations of motion for the chiral superfields are
\beq\label{eqm}
\bar{D}^2(e^{-gV}\bar{\Phi}^a_1 e^{gV})= -ih \Phi^b_2\Phi^c_3~[q(abc)-\bar{q}(acb)]
\eeq
and cyclic, where $(abc)\equiv \mbox{Tr}(T^aT^bT^c)$.

At the quantum level the theory is superconformal invariant
(and then finite) up to two loops if the coupling constants satisfy the following condition
(vanishing of the beta functions) \cite{FG,PSZ4}
\beq
|h|^2 \left[ 1 - \frac{1}{N^2}\left| q - \bar{q} \right|^2 \right] = g^2
\label{superconformal}
\eeq
Superconformal invariance at three loops has been discussed in \cite{RSS} for any $N$.
In the large $N$ limit this condition reduces simply to $|h|^2=g^2$, independently of
the value of $q$. In \cite{MPSZ} it has been proven that this is the {\em exact}
superconformal invariance condition for the large $N$ theory dual to the Lunin--Maldacena
supergravity background \cite{LM}.

In this paper we consider the ${\cal N}=1$ superconformal theory at {\em finite} $N$,
perturbatively defined by the condition (\ref{superconformal}) and investigate at the
quantum level some sectors of its chiral ring.

\section{The chiral ring of the $\beta$-deformed theory}\label{conj}

We are interested in studying perturbatively the structure of the chiral ring for
the $\beta$--deformed theory (\ref{baction}). As discussed in \cite{CDSW},
for a generic ${\cal N}=1$ SYM theory scalar operators in the chiral ring can be constructed
as products of scalar chiral superfields $\Phi_i$ and/or times $(W^{\a}W_{\a})$, where $W_\alpha$ is the chiral
field strength. In this paper we will focus only on the $\Phi$--sector, neglecting
operators with a dependence on $W_\a$.

In \cite{BL,BJL,LM} the single--trace sector of the chiral ring has been identified
as given by chiral operators of the form
${\rm Tr}(\Phi_1^{J_1}\Phi_2^{J_2}\Phi_3^{J_3})$ with weight $\Delta_0 = J_1+J_2+J_3$
and $(J_1, J_2, J_3) = (J,0,0), (0,J,0), (0,0,J), (J,J,J)$.
In \cite{FG, PSZ4} it has been shown perturbatively that
also the assignements $(J_1, J_2, J_3) = (1,1,0), (1,0,1), (0,1,1)$ give protected
operators.

This classification identifies the CPO's according to their dimension and
their charges with respect to the two $U(1)$ global invariances of the theory.
However, it does not give any information on
the precise form of the protected operator corresponding to a given set $(J_1, J_2, J_3)$,
which turns out to be in general a linear combination of single--trace operators with
different order of the fields inside the trace. Moreover, if we work at finite $N$, mixing
with multi--trace operators is also allowed.

A first example has been studied in \cite{FG} for the weight--3 sector.
There, it has been shown that the correct
expression for the protected operator correponding to $(J_1, J_2, J_3) = (1,1,1)$
is a linear combination
\beq
{\rm Tr}(\Phi_1\Phi_2\Phi_3)  + \alpha {\rm Tr}(\Phi_1\Phi_3\Phi_2)
\label{fg}
\eeq
where at one--loop
\beq
\alpha = \frac{(N^2-2) \bar{q}^2 + 2}{N^2-2 + 2\bar{q}^2}
\label{alpha}
\eeq
showing an explicit dependence on the coupling $\b$.

We are interested in the generalization of this result to higher loops in order to investigate
whether and how the linear combination gets modified order by order. Moreover, we
extend this analysis to other sectors of the chiral ring in order to discuss mixing
at finite $N$.

In general, given a set of primary operators ${\cal O}_i$
with the same dimension $\Delta_0$ and the same
global charges, we can read their anomalous dimensions perturbatively from the
matrix
of the two--point correlation functions. Precisely, this matrix has the form
\beq
\langle {\cal O}_i (x) {\cal O}_j (0) \rangle = \frac{1}{x^{2\Delta_0}} \left( A_{ij} -
\rho_{ij} \log{\mu^2 x^2} ~+~ \cdots ~\right)
\label{twopoint}
\eeq
where dots stay for higher powers in $\log{\mu^2 x^2}$.
Here $A$ is the mixing matrix, whereas $\rho$ signals the appearance of anomalous dimensions.
Both matrices are given as power series in the couplings.

In order to determine the anomalous dimensions we need diagonalize the two matrices by
performing the linear transformation ${\cal O}' = L {\cal O}$ which maps the operators
into an orthogonal basis of quasi--primaries.
In a perturbative approach it is easy to see \cite{APPSS,BRS}
that the diagonalization of the $\rho$ matrix  at order $n$ fixes the correct
orthogonalization (resolution of the mixing) at order $(n-1)$ uniquely, up to a residual
rotation among operators with the same anomalous dimension. This means that in general an
order $n$ calculation is required to determine the anomalous dimensions at this order
and the correct linear combinations of operators ${\cal O}_i$ at order $(n-1)$
which correspond to quasi--primaries with well--defined anomalous dimensions up to
order $n$.

In our case, since we are interested into {\bf chiral primary operators},
the procedure to determine perturbatively the
correct linear combination which corresponds to a protected operator is made simpler if we
also use the definition of chiral ring.

In our conventions the chiral ring is the set of chiral operators
which cannot be written, by using the equations of motion, as $\bar{D}^2 X$, being $X$ any
primary operator.

In general, given a set of linearly independent chiral operators ${\cal C}_i$, $i=1, \cdots,s$
characterized by the same classical scale dimension $\Delta_0$ and the same charges under the two
$U(1)$ flavor groups they will mix and we need solve the mixing in order to compute
their anomalous dimensions. Since we are working with chiral operators, we know a priori that
once we have orthogonalized as ${\cal C}'_i = L_{ij} {\cal C}_j$ in order to have well--defined
quasi--primary operators,
some of them will turn out to be descendant, i.e. they can
be written as $\bar{D}^2 X$ for some primary $X$.
The remaining operators will be necessarily primary chirals with vanishing anomalous
dimensions.

Exploiting this simple observation, in order to find the correct
expression for the protected operators, we then proceed as
follows: In a given $(J_1,J_2,J_3)$ sector, we first select all
the descendants, that is all the linear combinations \beq {\cal
D}_i = \sum_j d_j^{(i)} {\cal C}_j \label{combination} \eeq which
satisfy the condition \beq {\cal D}_i=\bar{D}^2 X_i
\label{descendant} \eeq Let us suppose that there are $i= 1,
\cdots , r \leq s $ independent linear combinations of this type.
Then, for a generic operator ${\cal P} = \sum_j c_j {\cal C}_j $
we impose the orthogonality condition \beq\label{cond} \langle
{\cal P} \bar{\mc D}_i\rangle=0\qquad i=1, \cdots ,r \eeq where
$\bar{\cal D}$ indicates the hermitian conjugate of ${\cal D}$.
These constraints provide $r$ equations for the $s$ unknowns
$c_j$. In this way we select a $(s-r)$-dimensional subspace of
operators orthogonal to the descendant ones. We can choose an
appropriate (orthogonal) basis in this subset, obtaining $(s-r)$
independent operators which are protected. This procedure has been
already applied in the undeformed ${\cal N}=4$ case \cite{EJSS}.

The problem of determining the CPO's of the theory is then traslated into the problem of
finding {\em all} the linear combinations of operators which satisfy the condition (\ref{descendant}).
In particular, since we are interested into a perturbative determination of the chiral
ring we need find descendants which solve eq. (\ref{descendant}) order by order in perturbation
theory. This can be done by introducing a perturbative definition of quantum chiral ring, as we
are now going to explain in detail.

\subsection{The perturbative quantum chiral ring}

As previously discussed, the chiral ring is defined as the set of chiral operators
orthogonal to null operators, i.e. linear combinations of chirals which can be written in
the form $\bar{D}^2 X$, $X$ primary. At the classical level a linear combination (\ref{combination})
gives rise to a null operator every time the coefficients $d_j^{(i)}$ are such that the operator
${\cal D}_i$
can be rewritten as a product of chiral superfields times $\frac{\delta W}{\delta \Phi_k}$,
where $W$ is the classical superpotential\footnote{This is true only for operators which are not
affected by Konishi--like anomalies or
as long as these anomalies do not enter the actual calculation (see the discussion at the beginning of Section 6).}
\beq
W = ih \left[ q ~{\rm Tr}(\Phi_1 \Phi_2 \Phi_3) - \bar{q}~{\rm Tr}(\Phi_1 \Phi_3 \Phi_2)
\right]
\label{superpot}
\eeq
Indeed, if this is the case, we can use the classical equations of motion
$\bar{D}^2 \bar{\Phi}_k = - \frac{\delta W}{\delta \Phi_k}$ to express the operator as in
(\ref{descendant}). It follows that we can alternatively define the chiral ring as
\beq
{\cal C} = \{ {\rm chiral ~op.'s}~ \mc P ~|~
\langle \mc P \bar{\mc D} \rangle = 0, ~{\rm for~ any}~
{\mc D} \sim ( ...\Phi .. \Phi .. \frac{\delta W}{\delta \Phi} ) \}
\label{classicalring}
\eeq
where in ${\mc D}$ we do not indicate trace structures
and flavor charges explicitly. In the undeformed ${\cal N}=4$ theory, an immediate consequence
of the definition (\ref{classicalring}) is that all the CPO's correspond to completly symmetric
representations of the $SU(3) \subset SU(4)$ R--symmetry group \cite{W}.

This definition for the chiral ring allows for a straightforward generalization at the quantum level.
Since the quantum dynamics of the elementary superfields is driven by the effective superpotential
rather than the classical $W$, it appears natural to define the quantum chiral ring as
\beq
{\cal C}_Q = \{ {\rm chiral ~op.'s} ~\mc P ~|~ \langle \mc P \bar{\mc D}_Q \rangle = 0,
~{\rm for~ any}~
{\mc D}_Q \sim ( ..\Phi ... \Phi ... \frac{\delta W_{eff}}{\delta \Phi} ) \}
\label{quantumring}
\eeq
where now $\mc D_Q$ is a {\em quantum} null operator. Using the quantum equations of motion
$\bar{D}^2 \frac{\delta K}{\delta \Phi_i} = -  \frac{\delta W_{eff}}{\delta \Phi_i}$ where $K$
is the effective K\"ahler potential which takes into account possible perturbative
D--term corrections, it is easy to see that $\mc D_Q$ is a null operator at the quantum level.
In the undeformed ${\cal N}=4$ case the symmetries of the theory constrain $\mc D_Q$ to be proportional
to $\mc D$ and the quantum chiral ring coincides with the classical one (\ref{classicalring}).

When $W_{eff}$ is determined perturbatively, eq.
(\ref{quantumring}) gives a perturbative definition of chiral
ring. Precisely, given $W_{eff}$ at a fixed perturbative
order\footnote{In principle, perturbative corrections to $W_{eff}$
would depend on both $g$ and $h$ couplings. Here we mean to use
the superconformal invariance condition to express $|h|^2$ as a
function of $g^2$ and write the perturbative expansion in powers
of the 't Hooft coupling $\l = \frac{g^2N}{4\pi^2}$.} \beq W_{eff}
=  W + \l W^{(1)}_{eff} + \l^2 W^{(2)}_{eff} + \cdots + \l^L
W^{(L)}_{eff} \eeq we can construct independent descendants
\footnote{As long as we are interested in orthogonalizing with
respect to the whole space generated by the descendants, we do not
need the precise form of pure descendants, but just a suitable set
of linear independent states. From now on we will refer to this
definition of quantum descendants.} at that order as \beq \mc D =
\mc D_0 + \l \mc D_1 + \l^2 \mc D_2 + \cdots + \l^L \mc D_L \qquad
, \qquad {\mc D}_i = \Phi\ldots\frac{\delta W_{eff}^{(i)}}{\delta
\Phi} \label{pertdesc} \eeq and determine the protected operators
$\mc P$ by imposing the orthogonality condition $\langle \mc P
\bar{\mc D} \rangle = 0$ order by order. Since $\mc P$ will be in
general a linear combination of single/multitrace operators, these
conditions allow to determine the coefficients of the linear
combination order by order in the couplings. If we set \beq \mc P
= \mc P_0 + \l \mc P_1 + \l^2 \mc P_2 + \cdots + \l^L \mc P_L \eeq
the perturbative corrections $\mc P_j$ will be determined by \bea
O(\l^0) &:&\qquad \langle {\mc P}_0 \bar{\mc D}_0 \rangle_0 =0 \nonumber\\
O(\l^1) &:& \qquad \langle {\mc P}_0 \bar{\mc D}_1 \rangle_0 +
\langle {\mc P}_0 \bar{\mc D}_0 \rangle_1 + \langle {\mc P}_1 \bar{\mc D}_0 \rangle_0 = 0
\label{pertcond} \\
\vdots~~~~  &~&\qquad\qquad\vdots \nonumber \\
O(\l^L) &:& \qquad \langle {\mc P}_0 \bar{\mc D}_L \rangle_0 +
\langle {\mc P}_0 \bar{\mc D}_{L-1} \rangle_1 + \cdots + \langle
{\mc P}_0 \bar{\mc D}_0 \rangle_L + \langle {\mc P}_1 \bar{\mc
D}_{L-1} \rangle_0 + \cdots + \langle {\mc P}_L \bar{\mc D}_0
\rangle_0 =0 \nonumber
\eea
where $\langle ~\rangle_j$ stands for the
two--point function at order $\l^j$.

Conditions (\ref{pertcond}) together with the general statement
that orthogonalization at order $(n-1)$ is sufficient for having
well--defined quasi--primary operators at order $n$, brings us to
formulate the following prescription: In order to determine
perturbatively the correct form of chiral operators with vanishing
anomalous dimension at order $n$ it is sufficient to determine the
effective superpotential at order $(n-1)$, select all the
descendant operators at that order by (\ref{pertdesc}) and impose the
conditions (\ref{pertcond}) up to order $(n-1)$.
In so doing, we gain a perturbative order at each step. Moreover,
in order to have all the descendants at a given order it is
sufficient to compute the effective superpotential once for all.

As follows from its definition, the structure of the chiral ring
is directly related to the structure of the effective
superpotential. Therefore, the perturbative corrections to the
CPO's depend on the perturbative corrections to the effective
superpotential. In particular, this explains universality
properties of the protected operators we will discuss in Section 5,
as for example the fact that in any case the
orthogonalization at tree level is sufficient for the protection
up to two loops.

\section{The effective superpotential at two--loops}

Since we are dealing with a superconformal (finite) theory any
correction to the effective action must be finite. By definition,
the effective superpotential corresponds to perturbative, finite
F--terms evaluated at zero momenta. It is given by {\em local}
contributions which are constrained by dimensions, $U(1) \times
U(1)$ flavor symmetry charges, reality and symmetry
(\ref{qsymmetry}) to have necessarily the form
\beq
W_{eff} = ih
\left[ b ~{\rm Tr}(\Phi_1 \Phi_2 \Phi_3) - \bar{b}~{\rm Tr}(\Phi_1
\Phi_3 \Phi_2) \right] ~+~ {\rm h.c.}
\label{Weff}
\eeq
The constant $b$ is given as an expansion in the couplings, $b = q(1 +
b_1 \l + b_2 \l^2 + \cdots) $, with coefficients $b_j$ which are
functions of $q$ and $N$, whereas $\bar{b}$ is the hermitian
conjugate. We note that in principle the symmetries of the theory
would only constrain the form of the superpotential to $W_{eff} =
\{ih \left[ b(q) ~{\rm Tr}(\Phi_1 \Phi_2 \Phi_3) +
b(-\bar{q})~{\rm Tr}(\Phi_1 \Phi_3 \Phi_2) \right] ~+~ {\rm
h.c.}\}$. However, it is easy to show that $ b(-\bar{q}) = -
\Bar{b(q)}$ since the $b_j$ coefficients are rational
functions of $q^2$ with real coefficients (loop diagrams always give real contributions
and they always contain an even number of extra chiral vertices compared to the tree--level
vertex).

At a given order $L$ we can have two kinds of corrections
to $W_{eff}$: Corrections which do not mix the two terms in the
superpotential and are then of the form \beq W_{eff}^{(L)} \sim
\l^L ~W \eeq where $W$ is the classical superpotential. These
contributions do not affect the structure of the descendant
operators at order $L$ since $\frac{\delta W_{eff}^{(L)}}{\delta
\Phi} \sim \frac{\delta W}{\delta \Phi}$ and $\mc D_L \sim \mc
D_0$. As a consequence at order $L$ the correlation function
$\langle \mc P_0 \bar{\mc D}_L \rangle_0$ in (\ref{pertcond})
vanishes and the protected operator is determined only by loop
corrections to its two--point function with descendants of lower
orders.

The second kind of corrections to $W_{eff}$ mixes the two terms in
$W$ and gives rise to a linear combination $W_{eff}^{(L)}$ of the
form (\ref{Weff}) which is not proportional to the classical
superpotential anymore. For these corrections the request for the
protected operator to be orthogonal to a descendant proportional
to $\frac{\delta W_{eff}^{(L)}}{\delta \Phi}$ modifies in general
its structure by contributions of order $\l^L$ proportional to
$\langle \mc P_0 \bar{\mc D}_L \rangle_0$.

In this Section we evaluate explicitly the effective superpotential up to two loops.
Our result is useful for determining the correct CPO's up to three loops.

\begin{figure} [ht]
\begin{center}
\begin{tabular}{cccc}
\epsfysize=3cm\epsfbox{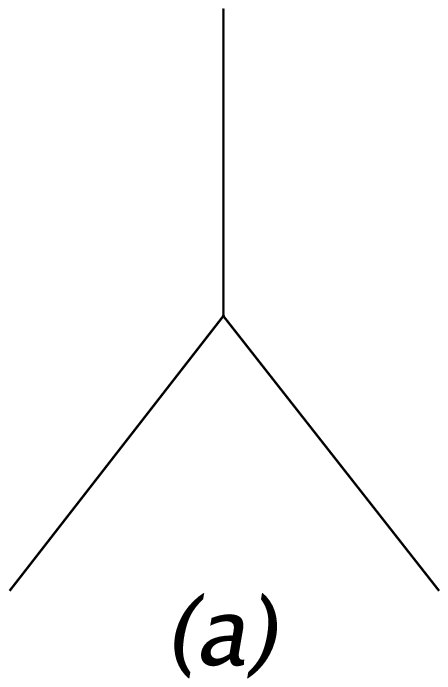} &
\epsfysize=3cm\epsfbox{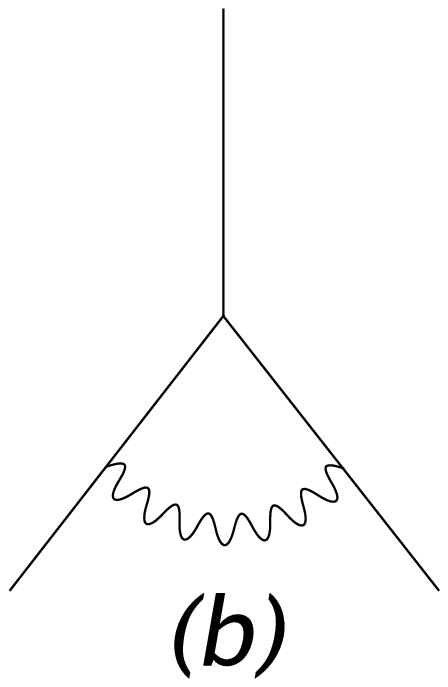} & \epsfysize=3cm\epsfbox{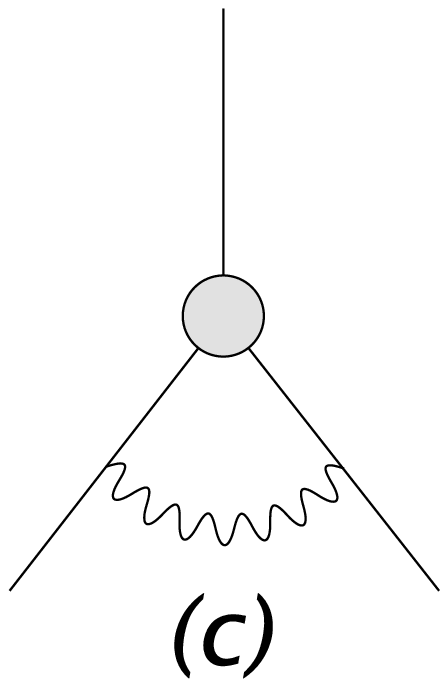} &
\epsfysize=3cm\epsfbox{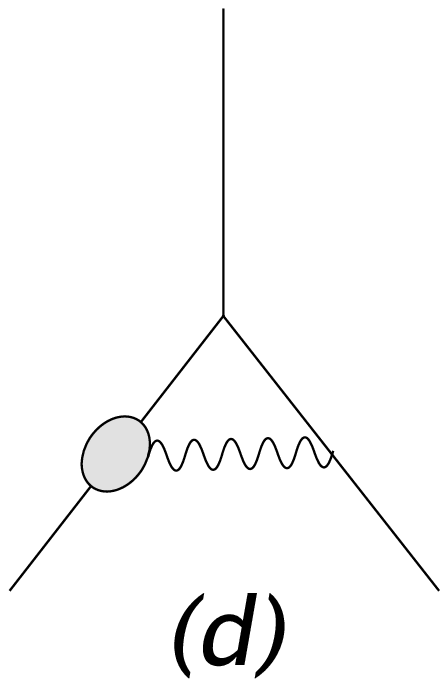} \\
~~ & ~~ & ~~ \\
\epsfysize=3cm\epsfbox{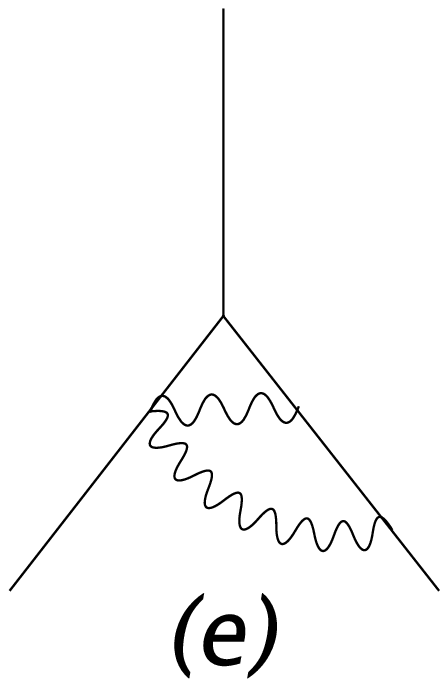} &
\epsfysize=3cm\epsfbox{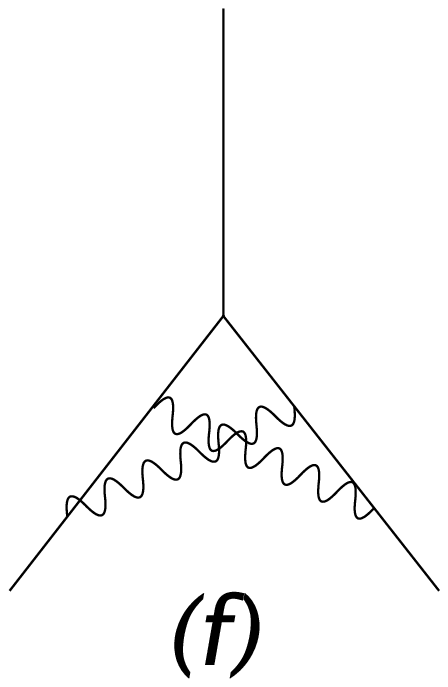} & \epsfysize=3cm\epsfbox{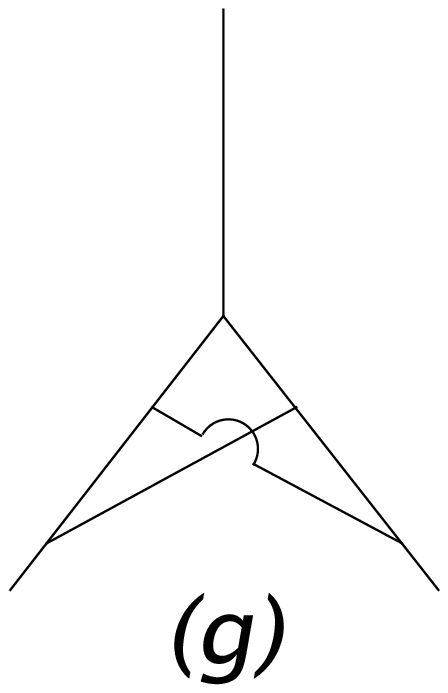}
\end{tabular}
\end{center}
\caption{Diagrams contributing to the effective superpotential up to two loops.}
\label{effsup}
\end{figure}

The diagrams contributing to the effective superpotential
up to this order are given in Fig. \ref{effsup} where the grey bullets indicate
the one--loop corrections to the chiral and gauge--chiral vertices, respectively.
These corrections are exactly the ones of the undeformed ${\cal N}=4$ theory
once we use the one--loop superconformal invariance condition (\ref{superconformal}).

The one--loop diagram \ref{effsup}b), compared with the tree level diagram \ref{effsup}a),
does not contain any extra $q$--deformed vertex. Moreover, using standard color identities
it is easy to see that its contribution is proportional to
$\l W$, where $W$ is the classical superpotential.

The same happens at two loops for the diagrams \ref{effsup}c), \ref{effsup}d) and
\ref{effsup}e)
which do not contain any extra $q$--deformed vertex and have a color structure
which does not mix the two traces, so reproducing $W$.

Diagram \ref{effsup}f) vanishes for color reasons.

Diagram \ref{effsup}g) contains four extra $q$--deformed vertices.
Moreover, by direct inspection one can
easily see that the nonplanar chiral structure which corrects the
tree level diagram mixes nontrivially the two terms of $W$. As a
result at two loops the superpotential undergoes a nontrivial
modification of the form
 \beq
W_{eff}^{(2)} \sim ih
\left[ q\, P~{\rm Tr}(\Phi_1 \Phi_2 \Phi_3)
- \bar{q} \, \bar{P}~ {\rm Tr}(\Phi_1 \Phi_3 \Phi_2) \right] ~+~ {\rm h.c.}
\eeq
with
\beq P
= \frac{(q^2 - 1)^3 [N^2 +3 + q^2 (3N^2 - 10 + 7 q^2)]}{q^2 [q^4 +1 + (N^2-2)q^2]^2}
\eeq
Here we have used $\bar{q} = 1/q$. We note that the nontrivial $q$--dependence
of this diagram is a direct consequence of its nonplanarity.
In fact,
as discussed in \cite{MPSZ} planar diagrams depend on the particular
combination $q\bar{q}=1$, while the nonplanar ones have generically nontrivial
phases. Moreover, a $q$--dependence has also been introduced
by using the superconformal condition (\ref{superconformal})
to express the coefficient $|h|^4$ from the four chiral vertices in
terms of $\l^2$.

To evaluate the various contributions from Fig. \ref{effsup} we
first perfom D--algebra to reduce superdiagrams to ordinary loop
diagrams and compute the corresponding integrals in momentum space
(for the description of the procedure and our conventions we refer
to \cite{PSZ1, PSZ2, PS, PSZ4}). As reported in Appendix A the one
and two--loop integrals are all finite and they give a
well--defined, local value for external momenta set to zero.
Therefore, collecting all the contributions, at two loops the
superpotential has the structure (\ref{Weff}) with
\beq
b = q \left[ (1 +
\l c_1 + \l^2 c_2) + \l^2 \frac{3}{8} \zeta(3) P \right]
\eeq
where the
coefficients $c_1,c_2$ are numbers, independent of $q$ and $N$,
determined by the loop integrals \ref{effsup}b) and \ref{effsup}c)--\ref{effsup}e),
respectively (we do not need their explicit values).

It follows that in general a descendant at this order will have the form
\beq
\mc D_Q = (1 + \l c_1 + \l^2 c_2) \mc D_0 + \l^2 \mc D_2
\label{descendant2}
\eeq
with $\mc D_2 \neq \mc D_0$.

\section{Chiral Primary Operators in the spin--$2$ sector}

\subsection{The $(J,1,0)$ flavor}

We start considering operators of the form ${\rm Tr}(\Phi_1^J
\Phi_2)$. In this case, due to the ciclicity of the trace, there
is no ambiguity in the ordering of the operators inside the trace.
In the large $N$ limit these operators do not belong to the chiral
ring, they are descendants and their anomalous dimensions have
been computed exactly \cite{MPSZ} for $J$ large.  However, for
finite $N$ they can mix with multitraces and give rise to linear
combinations of single and multi--trace operators which are
protected. We are going to construct them perturbatively up to
three loops. For simplicity we consider first the particular cases
of $J=3,4$ and postpone the discussion for generic $J$ at the end
of this Section.

\vskip 12pt \noindent \underline{The  $(3,1,0)$ case}: The first
nontrivial example where mixing conspires to give rise to
protected operators is for $J=3$. This sector contains the two
operators \beq \mc O_1=\mbox{Tr}(\Phi_1^3\Phi_2)\qquad , \qquad
\mc O_2=\mbox{Tr}(\Phi_1^2)\mbox{Tr}(\Phi_1\Phi_2) \label{delta4}
\eeq Using the classical equations of motion (\ref{eqm}), it is
easy to see that \beq \bar{D}^2\mbox{Tr}(\Phi_1^2 e^{-gV}
\bar{\Phi}_3 e^{gV})= \mbox{Tr}\left(\Phi_1^2 \frac{\d W}{\d
\Phi_3} \right) = -ih\left( q-\bar{q} \right)
[\mbox{Tr}(\Phi_1^3\Phi_2)-
\frac{1}{N}\mbox{Tr}(\Phi_1^2)\mbox{Tr}(\Phi_1\Phi_2)]
\label{desc1} \eeq and a descendant can be constructed as (we
always forget about the normalization of the operators) \beq \mc
D_0=\mc O_1-\frac{1}{N}\mc O_2 \label{desc2} \eeq The knowledge of
$\mc D_0$ allows us to determine the one--loop protected operator.
We consider the linear combination \beq \mc P_0=\mc
O_1+\alpha_0\,\mc O_2 \label{treeop} \eeq which, for any $\a_0
\neq -\frac{1}{N}$, gives an operator in the chiral ring. We then
impose the orthogonality condition $\langle\mc P_0\,\bar{\mc
D}_0\rangle_0 =  0$ and find
\beq \label{alpha02}
\a_0=-\frac{N^2-6}{2N}
\eeq
This result coincides with the one
found in \cite{RSS} where the one--loop CPO has been determined by
diagonalizing directly the one--loop anomalous dimension matrix.

In order to extend our analysis to higher loops we need establish
the correct form of the descendant operator order by order, as
described in Section 3. If we look at its perturbative definition
(\ref{pertdesc}) and the way the equations of motion work in this
case, we easily realize that as long as the effective
superpotential has the structure (\ref{Weff}) we obtain
\beq
{\rm Tr}\left(\Phi_1^2 \frac{\d W_{eff}}{\d \Phi_3}\right) =
-ih\left( b-\bar{b} \right) [\mbox{Tr}(\Phi_1^3\Phi_2)-
\frac{1}{N}\mbox{Tr}(\Phi_1^2)\mbox{Tr}(\Phi_1\Phi_2)]
\eeq
whatever $b$ might be (determined perturbatively at a given order). It
follows that the linear combination on the r.h.s. of this equation,
which is nothing but the operator (\ref{desc2}), is always a
descendant operator independently of the order we have computed
the coefficient $b$. Therefore we
conclude that (\ref{desc2}) is the {\em exact} quantum descendant
up to an overall coupling--dependent normalization factor,
that is $\mc D_Q \sim \mc D_0$.

An alternative way \cite{EJSS} to establish the relation $\mc
D_Q \sim \mc D_0$ is to consider the combination
\beq
\bar{D}^2\mbox{Tr}(\Phi_1^2 e^{-gV} \bar{\Phi}_3 e^{gV})+
ih\left( q-\bar{q} \right) [\mbox{Tr}(\Phi_1^3\Phi_2)-
\frac{1}{N}\mbox{Tr}(\Phi_1^2)\mbox{Tr}(\Phi_1\Phi_2)]
\label{linearcomb}
\eeq
which is zero at tree level
and check that it is order by order orthogonal to the three monomials
$\bar{D}^2\mbox{Tr}(\Phi_1^2 e^{-gV} \bar{\Phi}_3 e^{gV})$,
$\mbox{Tr}(\Phi_1^3\Phi_2)$ and
$\mbox{Tr}(\Phi_1^2)\mbox{Tr}(\Phi_1\Phi_2)$, separately. In fact, if this
is the case, there is no extra mixing of the linear combination
(\ref{linearcomb}) with the three operators at the quantum level
and (\ref{linearcomb}) must be necessarily zero at any order in perturbation
theory.
We have checked the absence of mixing perturbatively up to two loops
confirming our conclusion.

In order to determine the protected operator we consider the
linear combination \beq \mc P=\mc O_1+\alpha\,\mc O_2 \eeq with
$\a$ given as an expansion in $\l$ \beq \alpha=\alpha_0+\alpha_1
\, \l +\alpha_2\, \l^2 \,+O(\l^3) \label{alphaexp} \eeq In the
notation of Section 3 we have $\mc P_0 = \mc O_1 + \a_0 \mc O_2$
with $\a_0$ already determined in (\ref{alpha02}) and $\mc P_j =
\a_j \mc O_2$.

As a consequence of the relation $\mc D_Q \sim \mc D_0$ the
orthogonality conditions (\ref{pertcond}) become
\bea
&& O(\l)~ : \quad \quad \langle \mc P_0 \bar{\mc D}_{0} \rangle_{1} +
\langle \mc P_1 \bar{\mc D}_{0} \rangle_{0}= 0
\label{pertcond2}\\
&& O(\l^2): \quad \quad \langle \mc P_0 \bar{\mc D}_{0}
\rangle_{2} + \langle \mc P_1 \bar{\mc D}_{0} \rangle_{1} +\langle \mc P_2 \bar{\mc D}_{0} \rangle_{0}= 0
\label{pertcond3}
\eea
The first condition (\ref{pertcond2}) gives
\beq \a_1 = -
\frac{\langle (\mc O_1 + \a_0 \mc O_2) \bar{\mc D}_0 \rangle_1
}{\langle \mc O_2 \bar{\mc D}_0 \rangle_0 }
\label{computingalpha1}
\eeq
In order to select the diagrams which
contribute to the two--point function at the numerator we note
that the tree level correlation function at the denominator, when
computed in momentum space and in dimensional regularization ($n =
4 - 2\e$), is $1/\e$ divergent. This divergence signals the
well--known short distance singularity of any two--point function
of a conformal field theory.

\begin{figure}[ht]
\begin{center}
\begin{tabular}{c}
\epsfysize=3cm\epsfbox{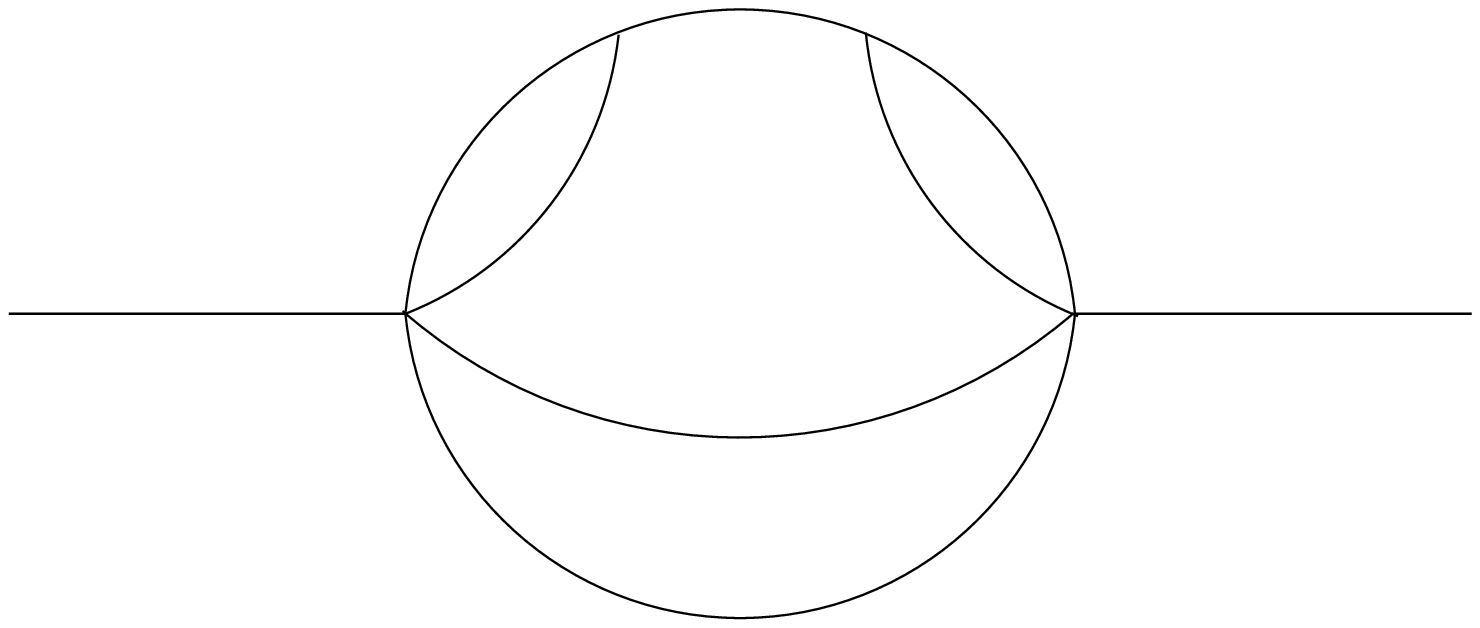}
\end{tabular}
\end{center}
\caption{One--loop diagram contributing to the evaluation of $\a_1$.}
\label{ear}
\end{figure}

If the denominator of (\ref{computingalpha1}) goes as $1/\e$,
in the numerator we can consider only divergent diagrams (finite
diagrams would not contribute in the $\e \to 0$ limit).
It is easy to show that
at this order the only diagram which we need take into account is
the one in Fig. \ref{ear} where on the left hand side we have an
insertion of
the operator $(\mc O_1 + \a_0 \mc O_2)$ while on the right hand side we
have $\bar{\mc D}_0$.

By a direct calculation one realizes that if
$\a_0$ is chosen as in (\ref{alpha02}) this diagram vanishes.
The reason is very simple to understand: If we cut the diagram
vertically at the very right end, close to the $\bar{\cal D}_0$
vertex, from the calculation it comes out that the left part would be
nothing but a one--loop divergent contribution to the operator $(\mc
O_1 + \a_0 \mc O_2)$ which vanishes since $\a_0$ has been determined
just to give a protected (not renormalized) operator at one--loop.

From the one--loop constraint we then read $\a_1 = 0$ and the
expression (\ref{treeop}) with $\a_0$ as in (\ref{alpha02})
corresponds to the protected chiral operator up to two loops.

Next we analyze the constraint (\ref{pertcond3}). Setting $\mc P_1
=0$ there,  we obtain
\beq \label{alpha2}
\alpha_2=-\frac{\langle
(\mc O_1 +\alpha_0  \mc O_2)\,\bar{\mc D}_0 \rangle_2}{\langle \mc
O_2\,\bar{\mc D}_0\rangle_0}
\eeq
and consequently the exact
expression for the CPO up to three loops.

\begin{figure}[ht]
\begin{center}
\begin{tabular}{cc}
\epsfysize=3cm\epsfbox{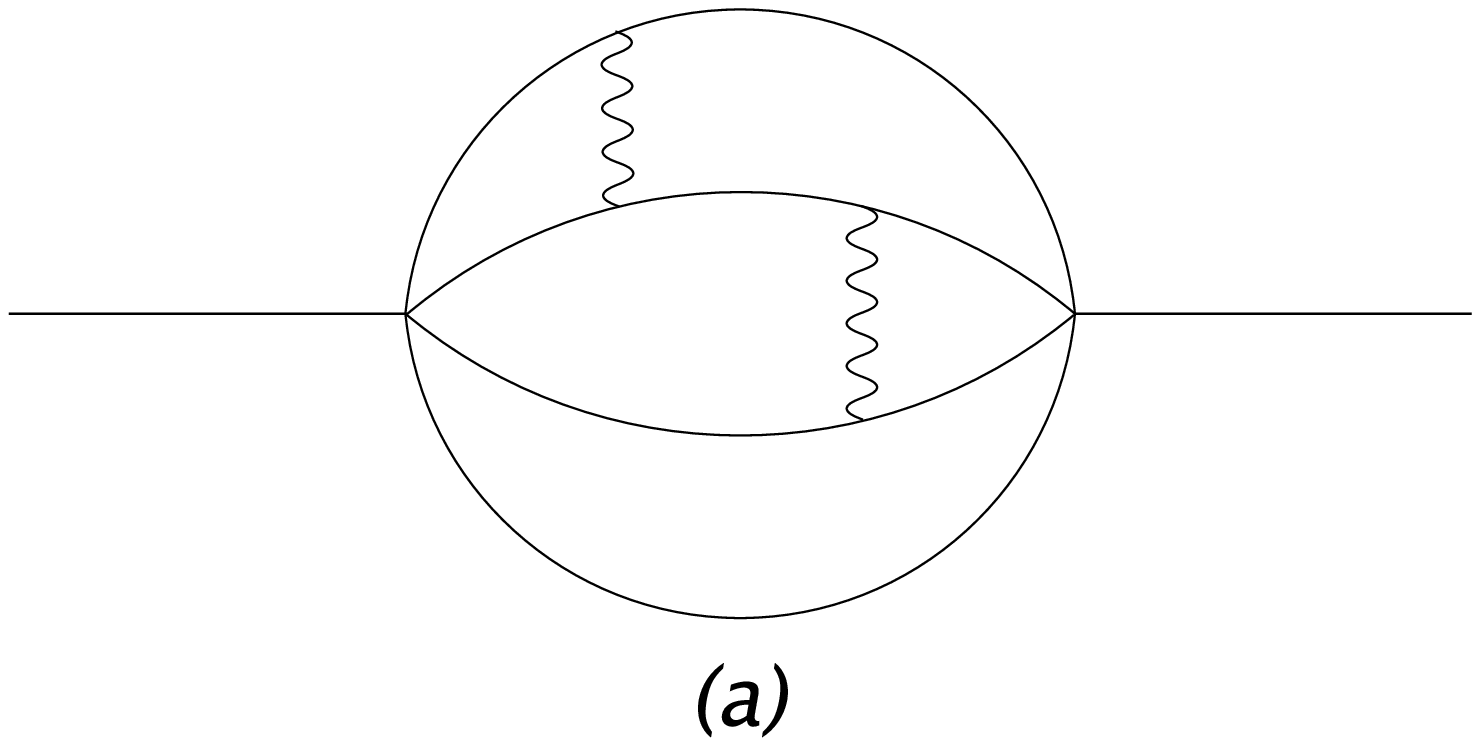} & \epsfysize=3cm\epsfbox{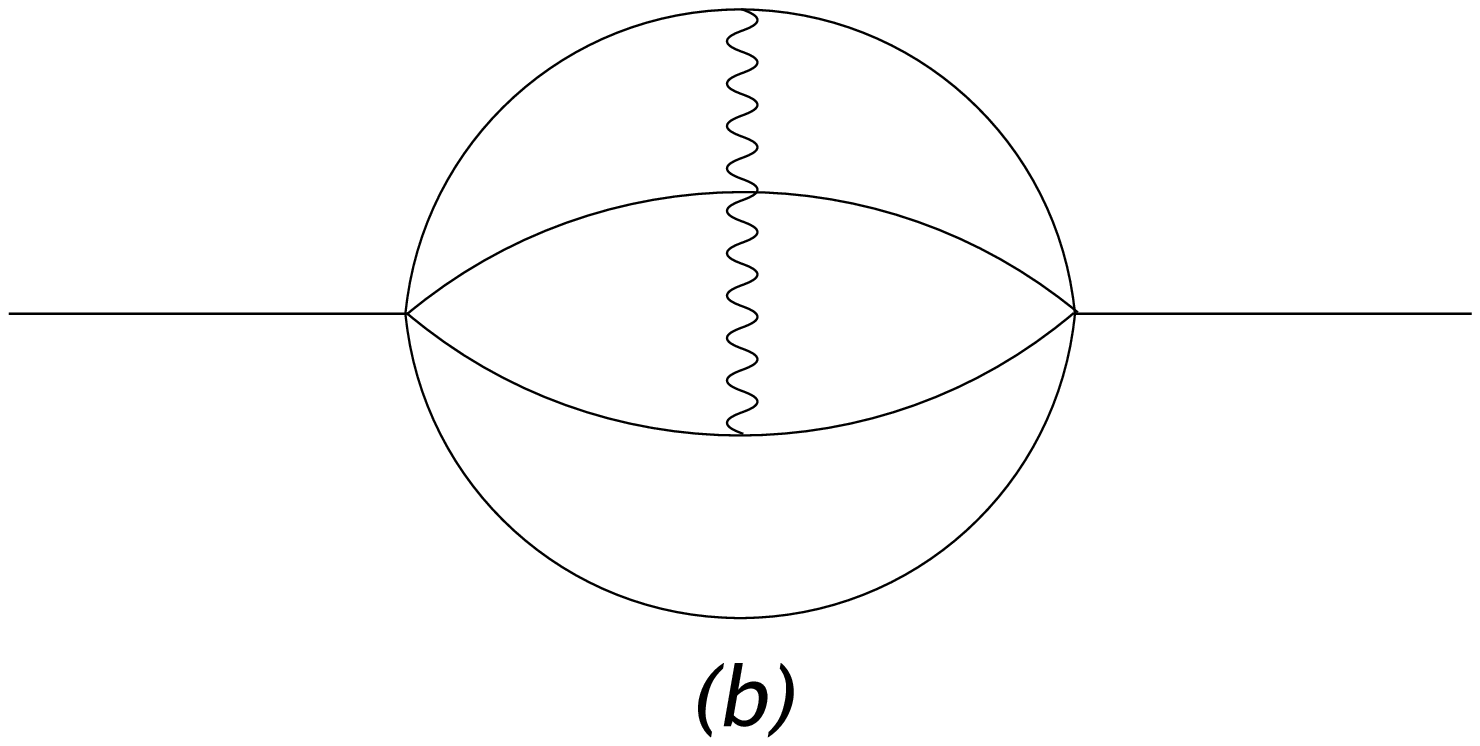} \\
\epsfysize=3cm\epsfbox{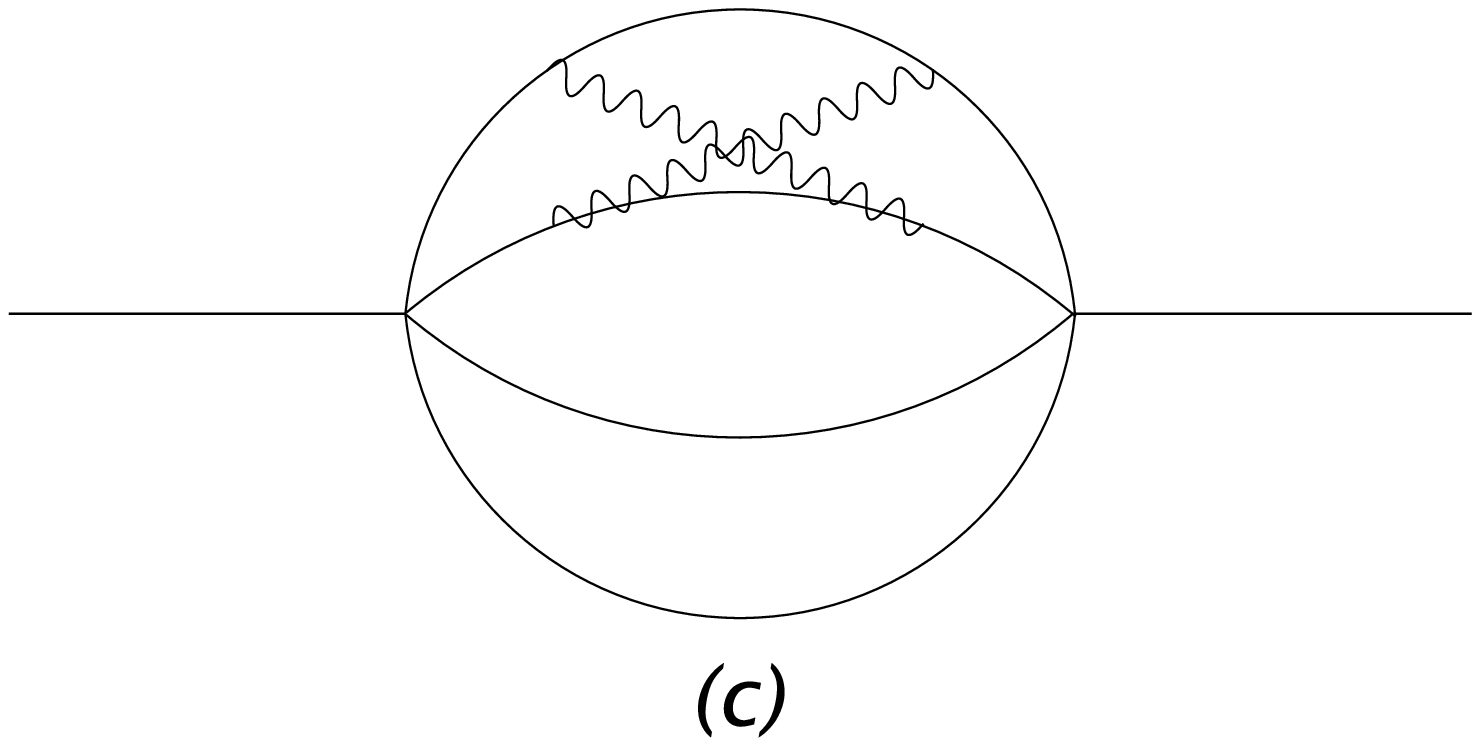} & \epsfysize=3cm\epsfbox{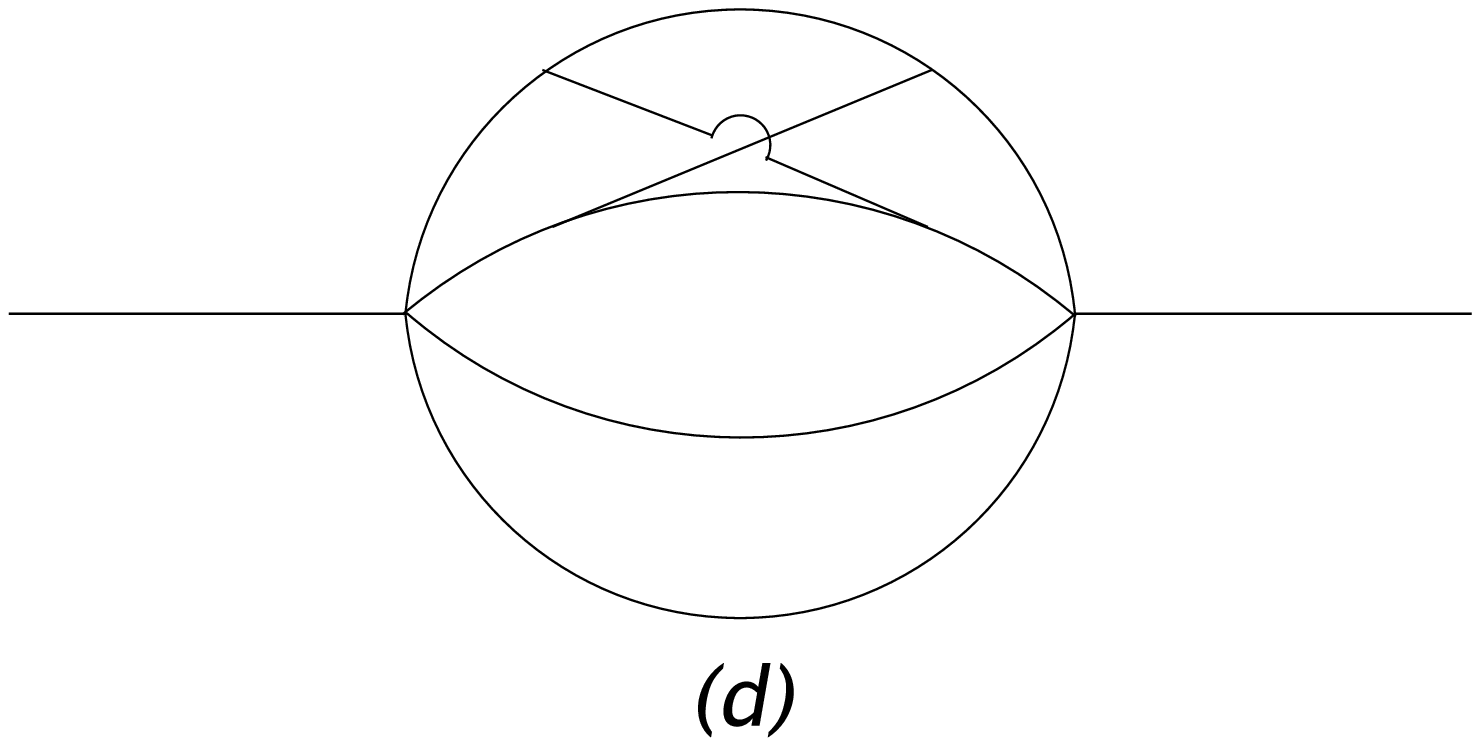}
\end{tabular}
\end{center}
\caption{Two-loop diagrams contributing to the evaluation of
$\alpha_2$.} \label{cross}
\end{figure}

Again we select only divergent diagrams contributing to the numerator.
They are given in Fig. \ref{cross}. We have not drawn diagrams
associated to the
two--loop anomalous dimension of the operator $(\mc O_1 + \a_0 \mc O_2)$
which vanish
when $\alpha_0$ is chosen as in (\ref{alpha02}).

\noindent
These diagrams contribute nontrivially to $\alpha_2$ since, cutting the
graphs at the very right hand side, their left parts cannot be recognized as
corrections to the tree--level operator (nontrivial mixing between $\mc O_1$
and $\mc O_2$ occurs). Evaluating the diagrams by using the results in Appendix A
we obtain
\beq \label{newalfa2}
\alpha_2=\frac{9(N^2-9)(q^2-1)^2[(N^4-8 N^2-8)(q^4+1) +2(N^4+8)q^2]}{80N [q^4+1 +(N^2-2)q^2]^2}
\, \zeta(3)
\eeq
where we have used the one--loop superconformal condition
(\ref{superconformal}) to express all the contributions of Fig. \ref{cross}
in terms of $\l^2$ and set $\bar{q}= 1/q$.

Therefore the protected operator $\mc P$ up to three--loops can be
written as
\beq
\mc P =\mc O_1-\frac{N^2-6}{2N}(1 +r \, \l^2 )\,
\mc O_2 \label{310protected}
\eeq
with
\beq
\label{esse}
r=\frac{\alpha_2}{\alpha_0}=-\frac{9(N^2-9)(q^2-1)^2[(N^4-8 N^2-8)(q^4+1)
+2(N^4+8)q^2]}{40(N^2-6)[q^4 +1 +(N^2-2)q^2]^2}\, \zeta(3)
\eeq
We note that in the 't Hooft limit, $N\rightarrow\infty$ and $\l$ fixed,
$\mc O_2$ dominates and gives the protected operator up to three loops. This is
consistent with the fact that, in the absence of mixing, the only
primary operators in a given $\Delta_0$ sector are necessarily
products of single--trace primaries ${\rm Tr}(\Phi_1^m)$ and ${\rm
Tr}(\Phi_1 \Phi_2)$.

\vskip 15pt
\noindent
\underline{The  $(4,1,0)$ case}: It is
interesting to analyze this case in detail since it is the first
case where more than one descendant appears.

This sector contains three independent operators
\beq
\mc O_1=\mbox{Tr}(\Phi_1^4\Phi_2)\quad , \quad \mc
O_2=\mbox{Tr}(\Phi_1^3)\mbox{Tr}(\Phi_1\Phi_2)\quad , \quad \mc
O_3=\mbox{Tr}(\Phi_1^2)\mbox{Tr}(\Phi_1^2\Phi_2) \label{delta5}
\eeq
Using the classical equations of motion (\ref{eqm}), we can write
\beq
\bar{D}^2\mbox{Tr}(\Phi_1^3 e^{-gV}\bar{\Phi}_3
e^{gV})= \mbox{Tr}\left(\Phi_1^3 \frac{\d W}{\d \Phi_3}\right) =
-ih\left( q-\bar{q} \right) [\mbox{Tr}(\Phi_1^4\Phi_2)-
\frac{1}{N}\mbox{Tr}(\Phi_1^3)\mbox{Tr}(\Phi_1\Phi_2)]
\eeq
\beq
\bar{D}^2\left[ \mbox{Tr}(\Phi_1^2)\mbox{Tr}(\Phi_1
e^{-gV}\bar{\Phi}_3e^{gV})\right]= \mbox{Tr}(\Phi_1^2)
\mbox{Tr}\left(\Phi_1 \frac{\d W}{\d \Phi_3}\right) = -ih\left(
q-\bar{q} \right) \mbox{Tr}(\Phi_1^2) \mbox{Tr}(\Phi_1^2\Phi_2)
\label{desc1b}
\eeq
Therefore, in this case we can consider the two descendants
\beq
\mc D_0^{(1)} = \mc O_1-\frac{1}{N}\mc O_2 \qquad , \qquad
\mc D_0^{(2)} = \mc O_3
\eeq
or any linear combination which
realizes an orthogonal basis in the subspace of weight--5
descendants.

As in the previous example it is easy to prove that, given the
particular structure (\ref{Weff}) of the effective superpotential
and the way the equations of motion enter the calculation, the
linear combinations $\mc D_0^{(1)}$ and $\mc D_0^{(2)}$ provide
two independent descendants even at the quantum level.

Proceeding as before we consider the linear combination
\beq \mc
P=\mc O_1+\alpha\,\mc O_2+\beta\,\mc O_3
\eeq
and choose the
constants $\alpha$ and $\beta$ (expanded in powers of $\l$) by
requiring $\mc P$ to be orthogonal to the two descendants up to
two loops.

Solving the constraints $\langle \mc P_0 \bar{\mc D}_0^{(i)}
\rangle_0$ at tree level we determine the correct expression for
the operator characterized by a vanishing one--loop anomalous
dimension \beq \mc P_0 = \mc O_1-\frac{N^2-12}{3N}\,\mc
O_2-\frac{2}{N}\,\mc O_3 \eeq As in the previous case, this
operator is automatically orthogonal to $\mc D^{(1)}_0$ and $\mc
D^{(2)}_0$ also at one loop and so we expect it to be protected up
to two loops.

The orthogonality at two loops can be
imposed exactly as in the previous case and allows to determine the
corrections $\a_2$ and $\b_2$. The diagrams contributing
are still the ones in Fig. \ref{cross} with one extra free chiral line running
between the two vertices.  Performing the calculation we find
the final expression for the operator protected up to three loops
\beq
\mc P=\mc O_1-\frac{N^2-12}{3N}\,(1+s_1 \, \l^2)\, \mc O_2
-\frac{2}{N}\,(1+ s_2 \, \l^2)\, \mc O_3
\eeq
where
\bea
s_1 = \frac{\alpha_2}{\a_0} &=&~  \frac{(N^2-16)(q^2-1)^2\,
[(11N^2 +21)(q^4+1)+2(N^2-21)q^2]}
{4(N^2-12)[q^4+1+(N^2-2)q^2]^2} \, \zeta(3) \non\\
s_2 = \frac{\beta_2}{\beta_0} &=& -\frac{(N^2-16)(q^2-1)^2[(N^2+5)(q^4+1)+2(N^2-5)q^2]}
{8[q^4+1 +(N^2-2)q^2]^2} \, \zeta(3)
\eea
Again, the
coefficients depend on $N$ in such a way that in the large $N$
limit only the ${\cal O}_2$ operator in (\ref{delta5}) survives in
agreement with the chiral ring content of the theory in the planar
limit.

We note that these coefficients, as well as $r$ in (\ref{esse})
are real. This is a consequence of the fact that in the sectors
studied so far the descendant operators are $q$--independent and
the two--point correlation functions are real.

\vskip 20pt
The previous analysis can be applied to the generic
operators of the form $(\Phi_1^J \Phi_2)$. The peculiar pattern $\mc
D_Q \sim \mc D_0$ for the descendants occurs in any $(J,1,0)$
sector since it only depends on the particular structure of the
superpotential and the particular way the equations of motion work
for this class of operators. Therefore, the determination of CPO's
proceeds as before. In particular, we expect the tree level orthogonality
condition to be still sufficient for protection up to two loops
since the only one--loop diagram relevant for the calculation would be
the vanishing one--loop anomalous dimension diagram in Fig. \ref{ear}.
At two loops diagrams of the kind drawn in Fig. \ref{cross} should be still the
only relevant ones.

Without entering the details of the calculations which would be
quite involved and not very illuminating, we can determine the
dimension of the corresponding chiral ring subspace, i.e. the
number of independent protected operators corresponding to $U(1)$
flavors $(J,1,0)$.

To be definite we consider $J$ even ($J=2p$). In this case the list of chirals we can construct is
\bea
&&  {\rm  ~single-trace}  ~\qquad {\rm Tr}(\Phi_1^{2p} \Phi_2)
\nonumber \\
&& {\rm ~ double-trace}  \qquad {\rm Tr}(\Phi_1^{m_1}) ~{\rm Tr}(\Phi_1^{2p-m_1} \Phi_2)
\qquad \qquad \qquad~ m_1 = 2, \cdots, 2p-1
\nonumber \\
&& {\rm  ~triple-trace}  ~~\qquad {\rm Tr}(\Phi_1^{m_1}) ~{\rm Tr}(\Phi_1^{m_2})
~{\rm Tr}(\Phi_1^{2p-m_1-m_2} \Phi_2)
\nonumber \\
&& \qquad \qquad \qquad   \qquad  \qquad \qquad
\qquad m_1 = 2, \cdots, p-1 , ~~~m_2 = m_1, \cdots, 2p-1-m_1
\nonumber \\
&& \qquad \qquad \vdots
\nonumber \\
&& ~~~~~p {\rm -trace}  \qquad \qquad \!\!\! {\rm Tr}(\Phi_1^2) \cdots
{\rm Tr}(\Phi_1^2)~{\rm Tr}(\Phi_1^2 \Phi_2) \,\,\, , \,\,
{\rm Tr}(\Phi_1^3)~{\rm Tr}(\Phi_1^2)\cdots {\rm Tr}(\Phi_1^2)~{\rm Tr}(\Phi_1 \Phi_2) \nonumber \\
\label{deltaJ}
\eea
In order to find how many independent primaries we can construct out of
(\ref{deltaJ}) we need first count how many descendants of the form (\ref{descendant})
we have. As explained in the previous
simple examples, given the generic $n$--trace, $\Delta_0=J$ sector, null conditions come from
considering the operators
\beq
{\rm Tr}(\Phi_1^{m_1})  \cdots {\rm Tr}(\Phi_1^{m_{n-1}}) \bar{D}^2 {\rm Tr}(\Phi_1^{2p-1-m_1 -
... - m_{n-1}} e^{-gV} \bar{\Phi}_3 e^{gV})
\label{null}
\eeq
as long as $2p- 1- m_1 - ... - m_{n-1}  \geq 1$. In fact, once we act with $\bar{D}^2$ on
$\bar{\Phi}_3$ and use the equations of motion (\ref{eqm}) we generate the linear combination
\bea
&& {\rm Tr}(\Phi_1^{m_1})  \cdots {\rm Tr}(\Phi_1^{m_{n-1}}) {\rm Tr}(\Phi_1^{2p-m_1 -
... - m_{n-1}} \Phi_2)
\nonumber \\
&&~~~~~~~~~~~~- \frac{1}{N} {\rm Tr}(\Phi_1^{m_1})  \cdots {\rm Tr}(\Phi_1^{m_{n-1}})
{\rm Tr}(\Phi_1^{2p-1-m_1 - ... - m_{n-1}}){\rm Tr}(\Phi_1 \Phi_2)
\eea
which is then a descendant. Therefore, the complete list of descendants is
\bea\label{listdesc}
&&  {\rm  ~single-trace}  ~\qquad \bar{D}^2 \;{\rm Tr}(\Phi_1^{2p-1} e^{-gV}\bar{\Phi}_3e^{gV})
\nonumber \\
&& {\rm ~ double-trace}  \qquad \bar{D}^2 \left[{\rm Tr}(\Phi_1^{m_1}) ~{\rm Tr}(\Phi_1^{2p-1-m_1}
e^{-gV}\bar{\Phi}_3e^{gV})\right] \qquad \qquad  m_1 = 2, \cdots, 2p-2
\nonumber \\
&& {\rm  ~triple-trace}  ~~\qquad \bar{D}^2 \left[{\rm Tr}(\Phi_1^{m_1}) ~{\rm Tr}(\Phi_1^{m_2})
~{\rm Tr}(\Phi_1^{2p-1-m_1-m_2}e^{-gV}\bar{\Phi}_3e^{gV})\right]
\nonumber \\
&& \qquad \qquad \qquad   \qquad  \qquad \qquad
\qquad m_1 = 2, \cdots, p-1, ~~~m_2 = m_1, \cdots, 2p-2-m_1
\nonumber \\
&& \qquad \qquad \vdots
\nonumber \\
&& ~~~~~~p{\rm -trace}  \qquad \quad ~\bar{D}^2 \left[{\rm Tr}(\Phi_1^2) \cdots
~{\rm Tr}(\Phi_1^2)~{\rm Tr}(\Phi_1 e^{-gV}\bar{\Phi}_3e^{gV} )\right]
\eea
Counting how many operators we have in (\ref{deltaJ}) and subtracting the number of descendants in
(\ref{listdesc}) we find that the number of protected chiral operators is $\sum_{n=2}^{p} X_n$
where $X_n$ is the number of partitions of $(2p-1)$ objects into $(n-1)$ boxes with
at least 2 objects per box.
Analogously, the number of chiral primary operators for $J$ odd is $\sum_{n=2}^{p+1} X_n$.

This result is consistent
with the number of primary operators which survive in the large $N$ limit where mixing
effects disappear and the chiral ring reduces to products of single--trace operators
${\rm Tr}(\Phi_1^k)$, ${\rm Tr}(\Phi_1 \Phi_2)$.

\subsection{The $(2,2,0)$ flavor}

In the class of more general operators with weights $(J_1,J_2,0)$ we consider the
particular case $J_1 = J_2 = 2$. This sector contains four operators, two single-- and two double--traces
\bea
&& \mc O_1 = \mbox{Tr}(\Phi_1^2\Phi_2^2) \qquad \qquad , \qquad
\mc O_2 =\mbox{Tr}(\Phi_1\Phi_2\Phi_1\Phi_2) \non\\
&& \mc O_3=\mbox{Tr}(\Phi_1^2)\mbox{Tr}(\Phi_2^2) \qquad , \qquad
\mc O_4 = \mbox{Tr}(\Phi_1\Phi_2)\mbox{Tr}(\Phi_1\Phi_2)
\eea
Using the classical equations of motion (\ref{eqm}), we can write
\bea
\bar{D}^2 \left[ \mbox{Tr}(\Phi_1\Phi_2 e^{-gV} \bar{\Phi}_3 e^{gV}) - 
\mbox{Tr}(\Phi_2\Phi_1 e^{-gV} \bar{\Phi}_3 e^{gV} ) \right]
&=&-ih(q+\bar{q})[\mc O_2-\mc O_1] \non\\
\bar{D}^2 \left[ \mbox{Tr}(\Phi_1\Phi_2 e^{-gV} \bar{\Phi}_3 e^{gV} ) + 
\mbox{Tr}(\Phi_2\Phi_1 e^{-gV} \bar{\Phi}_3 e^{gV} ) \right]
&=&-ih(q-\bar{q})[\mc O_1+\mc O_2-\frac{2}{N}\mc O_4] \non \\
\eea
We note that on the right hand side of these equations the $q$--dependence is still factored out
as it happened in the previous cases (see eqs. (\ref{desc1}, \ref{desc1b})).
Therefore, tree level descendants can be defined as linear combinations
\bea
\mc D^{(1)}_0 &=&\mc O_2-\mc O_1 \non\\
\mc D_0^{(2)}&=&\mc O_1+\mc O_2-\frac{2}{N}\mc O_4 \eea Because of
their $q$--independence these operators correspond indeed to
a suitable choice of quantum descendants.

The general structure of a chiral primary operator in this sector is
\beq
\mc P=\a \mc O_1+\b \,\mc O_2+\g \,\mc O_3+ \d \,\mc O_4
\eeq
where the coefficients are determined order by order by the orthogonality conditions
$\langle \mc P \bar{\mc D}^{(1)}_0 \rangle$ and $\langle \mc P \bar{\mc D}^{(2)}_0 \rangle$.
Having two conditions
for four unknowns we expect to single out two protected operators.

At tree level, for the particular choice $\a_0=2, \b_0 = 1$ and
$\a_0=1, \b_0=-1$, we find
\bea
&& \mc P^{(1)}=2\mc O_1+\mc O_2-\frac{N^2-6}{2N}(\mc O_3+2\mc O_4) \non \\
&& \mc P^{(2)}=\mc O_1-\mc O_2-\frac{N}{4}\mc O_3+N \mc O_4
\label{220protected}
\eea
These are one--loop protected operators and coincide with the ones found in \cite{RSS}.
They are not orthogonal but a basis can be easily constructed by considering linear combinations.

According to the general pattern already discussed for the previous cases we expect the operators
(\ref{220protected}) to be protected up to two loops.
The condition for these operators to be protected up to three loops requires instead nontrivial
$\l^2$--corrections
to (\ref{220protected}) which can be determined by solving the orthogonality constraints at this order.
The diagrams contributing
nontrivially to the 2--point functions are still the ones in Fig. \ref{cross}.
Since the final expressions are quite unreadable, we find convenient to fix
$\a_2=\b_2=0$ for both the CPO's and we obtain
\bea
&& \mc P^{(1)}=2\mc O_1+\mc O_2-\frac{N^2-6}{2N}(1+t_1 \,\l^2 )\mc O_3-\frac{N^2-6}{N}(1+t_2\,\l^2)\mc O_4\non\\
&& \mc P^{(2)}=\mc O_1-\mc O_2-\frac{N}{4}(1+u_1 \,\l^2 )\mc O_3+N(1+u_2\,\l^2)\mc O_4\non\\
\eea
where
\bea
&& t_1=-\frac{9(N^2-9)(q^2-1)^2[(N^4-6 N^2-4)(q^4+1) +2(N^4-2 N^2+4)q^2]}{20(N^2-6)[q^4 +1 +(N^2-2)q^2]^2}\, \zeta(3)\non\\
&& t_2=\frac{9(N^2-9)(N^2+2)(q^2-1)^4}{10(N^2-6)[q^4 +1 +(N^2-2)q^2]^2}\, \zeta(3)
\eea
and
\bea
&& u_1=-\frac{9(q^2-1)^2[(N^6-9 N^4-16 N^2+18)(q^4+1) +2(N^6-14 N^4+34 N^2-18)q^2]}{20 N^2 [q^4 +1 +(N^2-2)q^2]^2}\, \zeta(3)\non\\
&& u_2=\frac{9(q^2-1)^2[(N^4-31 N^2-18)(q^4+1) -2(7 N^4-13 N^2-18)q^2]}{40 N^2 [q^4 +1 +(N^2-2)q^2]^2}\, \zeta(3)
\eea

\section{Chiral Primary Operators in the spin--$3$ sector}

This sector contains operators of the form $(\Phi_1^k \Phi_2^l
\Phi_3^m)$ with all possible trace structures.

The simplest case is for $k=l=m=1$ and involves the two weight--3
operators
\beq \label{FGoperators} \mc O_1 =
\mbox{Tr}(\Phi_1\Phi_2\Phi_3)\qquad , \qquad \mc O_2
=\mbox{Tr}(\Phi_1\Phi_3\Phi_2)
\eeq
As already mentioned, the
correct one--loop expression for the protected operator has been
determined in \cite{FG} by computing directly the anomalous
dimension at that order.  It turns out that the protected operator
is a linear combination of the two operators (\ref{FGoperators})
with coefficient $\alpha$ as in (\ref{alpha}). The result has been
confirmed in \cite{RSS} by using a simplified approach based on
the evaluation of the difference between the one--loop two--point
function of the deformed theory and the one for the ${\cal N}=4$
case. This approach is very convenient since it avoids computing
many graphs containing gauge vertices but, as recognized by the
authors, in this case it cannot be pushed beyond one loop.

Using our procedure, we can easily re-derive the Freedman--Gursoy
result by working at tree level and extend it to two--loops by
performing a one--loop calculation. The correct application of our
procedure beyond this order would require a substantial
modification in the definition of quantum chiral ring
(\ref{quantumring}) since in this sector descendants of
Konishi--like operators are present and the equations of motion
need be supplemented by the Konishi anomaly term. As a consequence
the corresponding chiral ring sector necessarily contains
operators depending on $W^\a W_\a$.

In fact, from the anomalous conservation equation for the Konishi
current we can write
\beq
\bar{D}^2\mbox{Tr}(e^{-gV}\bar{\Phi}_ie^{gV}\Phi_i)=
-3ih[q\,\mbox{Tr}(\Phi_1\Phi_2\Phi_3)-
\bar{q}\mbox{Tr}(\Phi_1\Phi_3\Phi_2)] + \frac{1}{32\pi^2}
\mbox{Tr}(W^\a W_\a)
\eeq
We remind that in our conventions $W_\a
= i\bar{D}^2 (e^{-gV}D_\a e^{gV})$ and it is at least of order
$g$. From the previous identity it follows that a descendant
operator has to be constructed out of the two operators
(\ref{FGoperators}) plus the anomaly term
\beq \label{desc} \mc
D_0= q\mc O_1-\bar{q} \mc O_2 + \frac{i}{96\pi^2 \, h}
\mbox{Tr}(W^\a W_\a) \label{D123}
\eeq
However, since the operator
$\mbox{Tr}(W^\a W_\a)$ is of order $g^2$ and has vanishing tree
level two--point function with $\mc O_1$ and $\mc O_2$ it does not
enter the orthogonality conditions at tree level and one--loop.
Therefore we can safely use our procedure to find CPO's up to two
loops forgetting about the anomaly.

Thus
we consider the linear combination
\beq
\mc P_0 = \mc O_1+ \a_0 \,\mc O_2
\label{fg1}
\eeq
for any value of $\a_0 \neq -\bar{q}^2$. In order
to determine the exact expression for the CPO at one--loop we need
impose the operator to be orthogonal to the descendant
(\ref{desc}) at tree level. A simple calculation proves that
$\langle \mc P_0 \bar{\mc D}_0 \rangle_0 = 0$ iff $\a_0$ is given
in (\ref{alpha}), in agreement with the result of \cite{FG}.

At one loop first we need determine the correct expression for the
descendant at this order. As it follows from the calculations of
Section 3 at one loop the effective superpotential is
proportional to the tree level $W$ and the corresponding descendant operator
is still proportional to $\mc D_0$ in eq. (\ref{D123}).
Given the generic linear combination $\mc P = \mc O_1+ (\a_0 +  \a_1 \l) \,\mc O_2$
we then impose the orthogonality condition up to order $\l$ to
uniquely determine $\a_1$ as in (\ref{computingalpha1}).
As in the previous examples, if $\a_0$ is given in (\ref{alpha})
the $\a_1$ coefficient is identically zero being this a consequence of the
one--loop protection of $\mc P_0$. Therefore the expression (\ref{fg1}) with
$\a_0$ given in (\ref{alpha}) corresponds to the protected chiral operator up
to two loops.

\vskip 15pt
The next case we investigate is for $k=2,\,l=m=1$. There are five
operators
\bea
\mc O_1=\mbox{Tr}(\Phi_1^2\Phi_2\Phi_3) \qquad &,&
\qquad \mc O_2 = \mbox{Tr}(\Phi_1^2\Phi_3\Phi_2) \qquad , \qquad
\mc O_3 = \mbox{Tr}(\Phi_1\Phi_2\Phi_1\Phi_3) \non\\
\mc O_4=\mbox{Tr}(\Phi_1^2)\mbox{Tr}(\Phi_2\Phi_3) \qquad &,&
\qquad \mc O_5 = \mbox{Tr}(\Phi_1\Phi_2)\mbox{Tr}(\Phi_1\Phi_3)
\eea Using the classical equations of motion (\ref{eqm}) we
can write three descendants
\bea\label{spin3desc}
\mc D^{(1)}_0 &=&q\mc O_3-\bar{q}\mc O_2-\frac{1}{N}(q-\bar{q})\mc O_5 \non\\
\mc D^{(2)}_0 &=&q\mc O_1-
\bar{q}\mc O_3-\frac{1}{N}(q-\bar{q})\mc O_5 \\
\mc D^{(3)}_0 &=&q\mc O_1-\bar{q}\mc O_2-\frac{1}{N}(q-\bar{q})\mc
O_4 \non
\eea
We expect to find out two protected operators of the form
\beq
\mc P=\a \,\mc O_1+\b \,\mc O_2+\g \,\mc O_3+\d \,\mc
O_4+\e \,\mc O_5 \label{CPO5}
\eeq
By imposing the tree-level
orthogonality condition with respect to the three $\mc D^{(i)}_0$
we can fix for instance $\g$, $\d$ and $\e$ in terms of
$\a$ and $\b$. The calculation proceeds exactly as in the
previous case and we find
\bea
\g&=&\frac{\a[q^4-2q^2+1-N^2]-\b[(1-N^2)q^4-2q^2+1]}{N^2(q^4-1)} \non\\
\d&=&\frac{\a[(N^2+2)q^4+2(N^2-2)q^2+N^4-5N^2+2]}{2N^3(q^4-1)} \non\\
&~~& -\frac{\b[(N^4-5N^2+2)q^4+2(N^2-2)q^2+N^2+2]}{2N^3(q^4-1)} \non\\
\e&=&\frac{\a[2(N^2+1)q^4+(N^4-4)q^2+N^4-4N^2+2]}{N^3(q^4-1)} \non\\
&~~&- \frac{\b[(N^4-4N^2+2)q^4+(N^4-4)q^2+2(N^2+1)]}{N^3(q^4-1)}
\label{CPOcoeffs}
\eea
We expect these operators to have a
vanishing anomalous dimension at one loop. If we set $\a=\b=1$
and $\a=-\b=1$, we recover the two protected operators found
in \cite{RSS}.

As in the previous cases, the operators $\mc D^{(1)}_0$, $\mc
D^{(2)}_0$ and $\mc D^{(3)}_0$ keep being good descendants at one loop.
Moreover, the one--loop orthogonality conditions do not modify the
CPO's (\ref{CPO5}, \ref{CPOcoeffs}) and we expect these operators
to have a vanishing two--loop anomalous dimension.

If we were to push our calculation beyond this order we should
first determine the descendant operators at two loops. It is easy
to realize that in this case the relation $\mc D_Q \sim \mc D_0$
does not hold anymore, for two simple reasons:

\noindent 1) At higher orders the Konishi anomaly cannot be
ignored anymore. In particular, the correct expression for the
descendant operators from two loops on will have a nontrivial
dependence on $(W^\a W_\a)$.

\noindent 2) Differently from the spin--2 case, the nontrivial
corrections to the effective superpotential which appear at two
loops determine nontrivial corrections to the descendants since in
this case they depend on $q$ not only through an overall
coefficient (see eq. (\ref{spin3desc})).

\section{The full Leigh--Strassler deformation}

From a field theory point of view it is interesting to investigate
the quantum properties of the full Leigh-Strassler $\mc N =1$
deformation of the $\mc N=4$ SYM theory given by the action
\cite{LS} \bea\label{action} S &=& \int
d^8z\mbox{Tr}(e^{-gV}\bar{\Phi}_ie^{gV}\Phi^i)+\frac{1}{2g^2}
\int d^6z\mbox{Tr}(W^{\alpha}W_{\alpha}) \non \\
&&~~ + \left\{ ih\int d^6 z \mbox{Tr}(q\,\Phi_1\Phi_2\Phi_3-\bar{q}\,\Phi_1\Phi_3\Phi_2)
+ \frac{ih'}{3}\int d^6 z \mbox{Tr}(\Phi_1^3+\Phi_2^3+\Phi_3^3) ~+~ {\rm h.c.} \right\} \non \\
\eea
The superpotential now breaks the original $SU(4)$ $R$--symmetry to $U(1)_R$
and no extra $U(1)$'s are left. However, the action is still invariant under the cyclic
permutation of $(\Phi_1, \Phi_2,\Phi_3)$ and the symmetry (\ref{qsymmetry}). Moreover, a
second $Z_3$ is left corresponding to
\beq
(\Phi_1,\Phi_2,\Phi_3) \quad \rightarrow \quad (\Phi_1,z\Phi_2,z^2\Phi_3) \non
\label{Z3}
\eeq
where $z$ is a cubic root of unity.

The equations of motion derived from (\ref{action}) are
\bea
\label{eom2}
\bar{D}^2(e^{-gV}\bar{\Phi}^a_1 e^{gV})&=&-ih\Phi_2^b\Phi_3^c\left[q(abc)-\bar{q}(acb)\right]-
ih'\Phi_1^b\Phi_1^c(abc) \non\\
\bar{D}^2(e^{-gV}\bar{\Phi}^b_2 e^{gV})&=&-ih\Phi_1^a\Phi_3^c\left[q(abc)-\bar{q}(acb)\right]-
ih'\Phi_2^a\Phi_2^c(abc) \\
\bar{D}^2(e^{-gV}\bar{\Phi}^c_3 e^{gV})&=&-ih\Phi_1^a\Phi_2^b\left[q(abc)-\bar{q}(acb)\right]-
ih'\Phi_3^a\Phi_3^b(abc) \non
\eea

As discussed in \cite{SV,LS} the request for the anomalous
dimensions of the elementary chiral superfields to vanish
guarantees the theory to be superconformal invariant. Since the
three chirals have the same anomalous dimension due to the cyclic
$Z_3$ symmetry, superconformal invariance requires a single
condition $\gamma(g,h,h',\beta)=0$ and we find a
three--dimensional complex manifold of fixed points.

In general we do not know the superconformal condition exactly.
However it is possible to perform a perturbative analysis and
define the superconformal theory order by order in the couplings.

To this purpose we evaluate the anomalous dimension of the chiral superfield $\Phi_i$ up to
two loops. The calculation can be carried on exactly as in the case of $h'=0$
by taking into account that compared to the previous case the present action contains
three extra chiral vertices of the form $\frac{ih'}{6} d_{abc} \Phi_i^a\Phi_i^b\Phi_i^c$, $i=1,2,3$.

As long as we deal with diagrams which do not contain the new $h'$ vertices we have exactly
the same contributions as in the $h'=0$ theory \cite{FG,PSZ4}. We only need evaluate
all the diagrams which contain these extra vertices.

At one loop, besides the $h$--chiral and the mixed gauge--chiral self--energy diagrams \cite{PSZ4} we
have a $h'$--chiral self--energy graph whose contribution is proportional to $|h'|^2$.
This new diagram modifies the one--loop superconformal condition (\ref{superconformal}) as
\beq \label{superconformal2}
\left[ |h|^2\left(1-\frac{1}{N^2}\left|q-\bar{q}\right|^2\right)+
|h'|^2\frac{N^2-4}{2 N^2}\right]=g^2
\eeq
in agreement with \cite{SR,OASR,NP}. As for the $h'=0$ case it is easy to verify that the one--loop condition
is sufficient to guarantee the vanishing of the beta functions (i.e. superconformal invariance)
up to two loops.

Once the theory is made finite we are interested in the perturbative evaluation of
{\em finite} corrections to the superpotential.
In this case the symmetries of the theory force the effective superpotential
to have the form
\beq
 W_{eff} = ih\int d^6 z \mbox{Tr}[b(q) \,\Phi_1\Phi_2\Phi_3 +
 b(-\bar{q})\,\Phi_1\Phi_3\Phi_2]
+ \frac{ih'}{3} d \int d^6 z \mbox{Tr}(\Phi_1^3+\Phi_2^3+\Phi_3^3) ~+~ {\rm h.c.}
\label{Wcorrected}
\eeq
where the coefficients $b$ and $d$ are determined as double power expansions in the couplings $h$
and $h'$ \footnote{Here we use the superconformal condition (\ref{superconformal2})
to express $g^2$ as a function of $h$ and $h'$. Any other choice would be equally acceptable.}.
In particular, the invariance under cyclic permutations of the superfields requires the $d$ correction
to be the same for the three $\Phi_i^3$ terms, whereas the other global symmetries
force the particular $q$ dependence of the corrections to $(\Phi_1\Phi_2\Phi_3)$ and
$(\Phi_1\Phi_3\Phi_2)$. We note that in this case we cannot apply the previous arguments
(see the discussion after eq. (\ref{Weff})) to
state that $b(-\bar{q}) = - \Bar{b(q)}$ since
the perturbative corrections to $(\Phi_1\Phi_2\Phi_3)$ and $(\Phi_1\Phi_3\Phi_2)$ are not
always proportional to $q$ times functions of $q^2$. In fact, it is still true that diagrams
contributing to the effective potential
contain an even number of extra chiral vertices compared to the tree level diagrams, but part of
these vertices could be $h'$--vertices not carrying any $q$--dependence.

The topologies of diagrams contributing to the superpotential up to two loops are still
the ones in Fig. \ref{effsup} where now chiral vertices may be either
$h$ or $h'$ vertices. Performing the explicit calculation as in Section 4 we discover that
at one loop the various terms in the superpotential do not mix and receive separate corrections
still proportional to the classical terms.
Precisely, we find that  $W^{(1)}_{eff}$ coincides with $W$, up to an overall constant coefficient.
This is also true at two loops for the diagrams \ref{effsup}c), \ref{effsup}d) and \ref{effsup}e),
whereas the diagram \ref{effsup}g) with all possible configurations of $h$ and $h'$ vertices
mixes nontrivially the various terms of the superpotential. Similarly to what happens for
the $\beta$--deformed
theory, this leads to a nontrivial correction $W_{eff}^{(2)}$ which has the form (\ref{Wcorrected})
but with the $b$ and $d$ coefficients nontrivially corrected by functions of $q$ and $N$.
We then expect descendant operators to get modified at this order as in the previous case (see discussion
around eq. (\ref{descendant2})).

The exact supergravity dual of the theory (\ref{action}) is still unknown even if few steps
towards it have been undertaken in \cite{GPPZAKY}.
However, it is interesting to investigate the nature of composite operators of the
superconformal field theory waiting for the discovery of the exact correspondence of these operators
to superstring states.

The chiral ring for the $h'$-deformed theory is not known in general (however, see \cite{BJL}).
Compared to the chiral ring
of the $\b$--deformed theory ($h'=0$) which contains operators of the form ${\rm Tr}(\Phi_i^J)$,
${\rm Tr}(\Phi_1^J\Phi_2^J\Phi_3^J)$ plus the particular operators ${\rm Tr}(\Phi_i\Phi_j)$,
$i \neq j$, we expect the chiral ring of the present theory to be more complicated because
of the lower number of global symmetries present.

Here we exploit the general procedure described in Section 3 to
move the first steps towards the determination of chiral primary
operators. In particular, we concentrate on the first simple cases
of matter chiral operators with dimensions $\Delta_0=2,3$ and
study how turning on the $h'$-interaction may affect their quantum
properties. We then take advantage of these results to make a
preliminary discussion of the CPO content for generic scale
dimensions.

\subsection{Chiral ring: The  $\D_0 = 2$ sector}

Weight--2 chiral operators are
$\mbox{Tr}(\Phi_i^2)$ and $\mbox{Tr}(\Phi_i \Phi_j)$, $i \neq j$.
These operators can be classified as in Table 1 according to their charge $\mc Q$ with respect to the
$Z_3$ symmetry (\ref{Z3}).

\begin{table}[ht]
\begin{center}
\begin{tabular}{|*{3}{c|}|}
\hline
$\mc Q = 0 $ & $\mc Q = 1$ & $\mc Q = 2$ \\
\hline
\hline
$\mc O_{11} =\mbox{Tr}(\Phi_1^2)~~~~$ & $\mc O_{33} =\mbox{Tr}(\Phi_3^2)~~~~$ & $\mc O_{22} =
\mbox{Tr}(\Phi_2^2)~~~~$ \\
\hline
$\mc O_{23}=\mbox{Tr}(\Phi_2 \Phi_3)$ & $\mc O_{12}=\mbox{Tr}(\Phi_1 \Phi_2)$ & $\mc O_{13}=
\mbox{Tr}(\Phi_1 \Phi_3)$ \\
\hline
\end{tabular}
\end{center}
\caption{Operators with $\Delta_0=2$.}
\label{opw2}
\end{table}

The charged sectors can be obtained from the $\mc Q = 0$ one by
successive applications of cyclic $Z_3$--permutations $\Phi_i
\rightarrow \Phi_{i+1}$. This is the reason why the three sectors
contain the same number of operators. In the $h'=0$ theory their
anomalous dimensions have been computed up to two loops and found
to be vanishing \cite{FG, PSZ4}. According to our discussion in
Section 3 this was an expected result since for these operators
there is no way to use the equations of motion (\ref{eqm}) to
write them as $\bar{D}^2 X$. Therefore they must be necessarily
primaries and belong to the classical chiral ring. Since this
sector does not contain descendants this property is mantained at
the quantum level. In the $h'=0$ case these operators have different
$U(1)$ flavor charges and do not mix. The matrix of their
two--point functions is then diagonal and receives finite
corrections at two loops \cite{PSZ4}.

The same analysis can be applied in the present case. Again, there
is no way to write these operators as descendants by using the
classical equations of motion (\ref{eom2}). Therefore, we expect
them to belong to the chiral ring.

In order to check that these operators do not get renormalized but their
correlators might receive finite corrections we compute directly
their two--point functions.

The smaller number of global symmetries surviving the
$h'$--deformation do not prevent the operators to mix. For
instance the operator $\mbox{Tr}(\Phi_1^2)$ can mix with
$\mbox{Tr}(\Phi_2 \Phi_3)$ since they have the same charge under
the $Z_3$ symmetry (\ref{Z3}). Therefore, we need compute the
non--diagonal matrix of their two--point functions.

To this purpose we concentrate on the operators $\mc O_{11}$ and  $\mc O_{23}$
and evaluate all the correlators up to two loops.
The calculation goes exactly as in the
$h'=0$ theory with the understanding of adding contributions from diagrams containing the
new $h'$--vertices.

At one--loop, as in the undeformed \cite{PSZ1,PSZ2} and the $\b$--deformed cases \cite{PSZ4}
we do not find any divergent nor finite contributions to the two--point functions
as long as the superconformal condition (\ref{superconformal2}) holds.

\begin{figure}[ht]
\begin{center}
\begin{tabular}{ccc}
\epsfysize=2.4cm\epsfbox{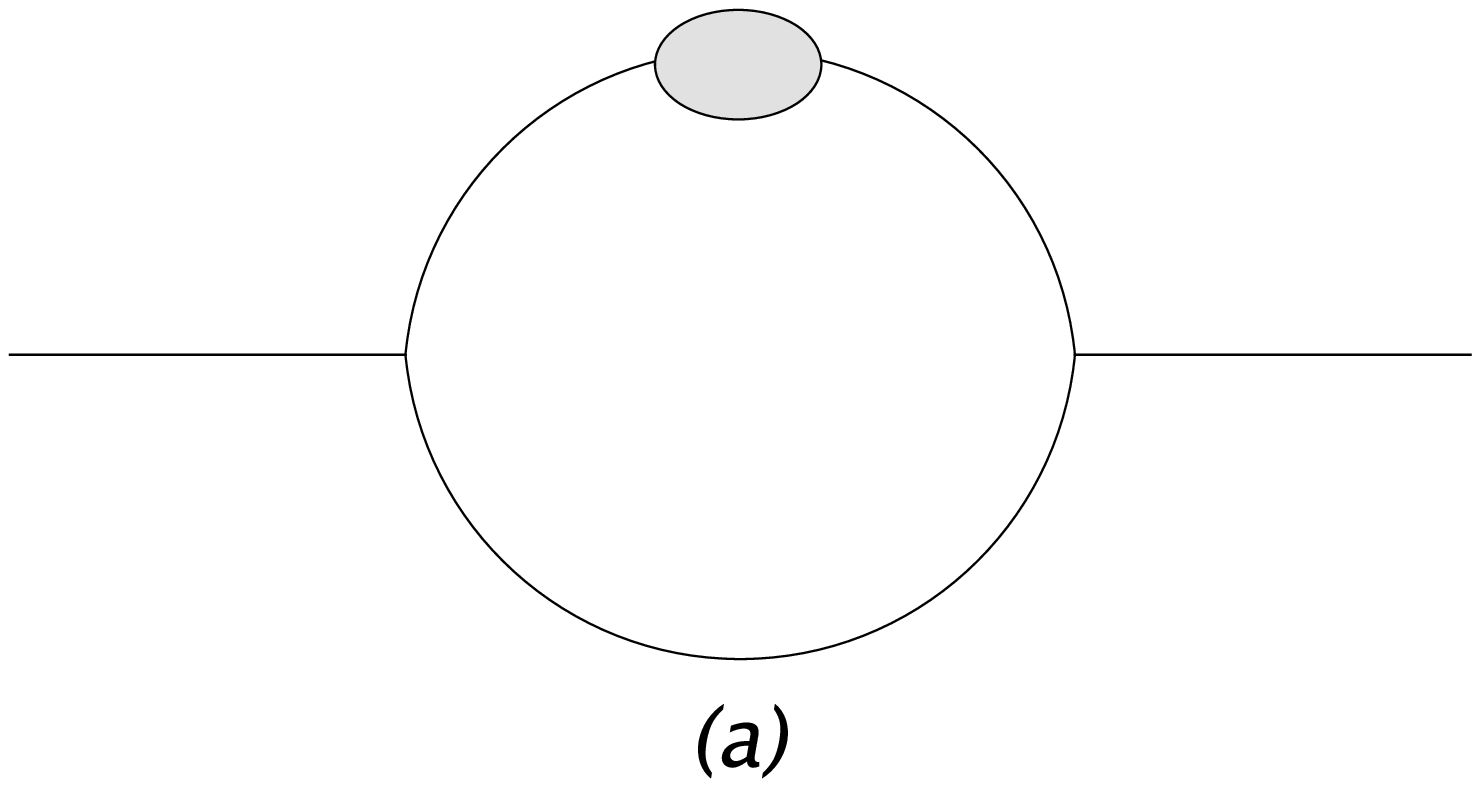} & \epsfysize=2.4cm\epsfbox{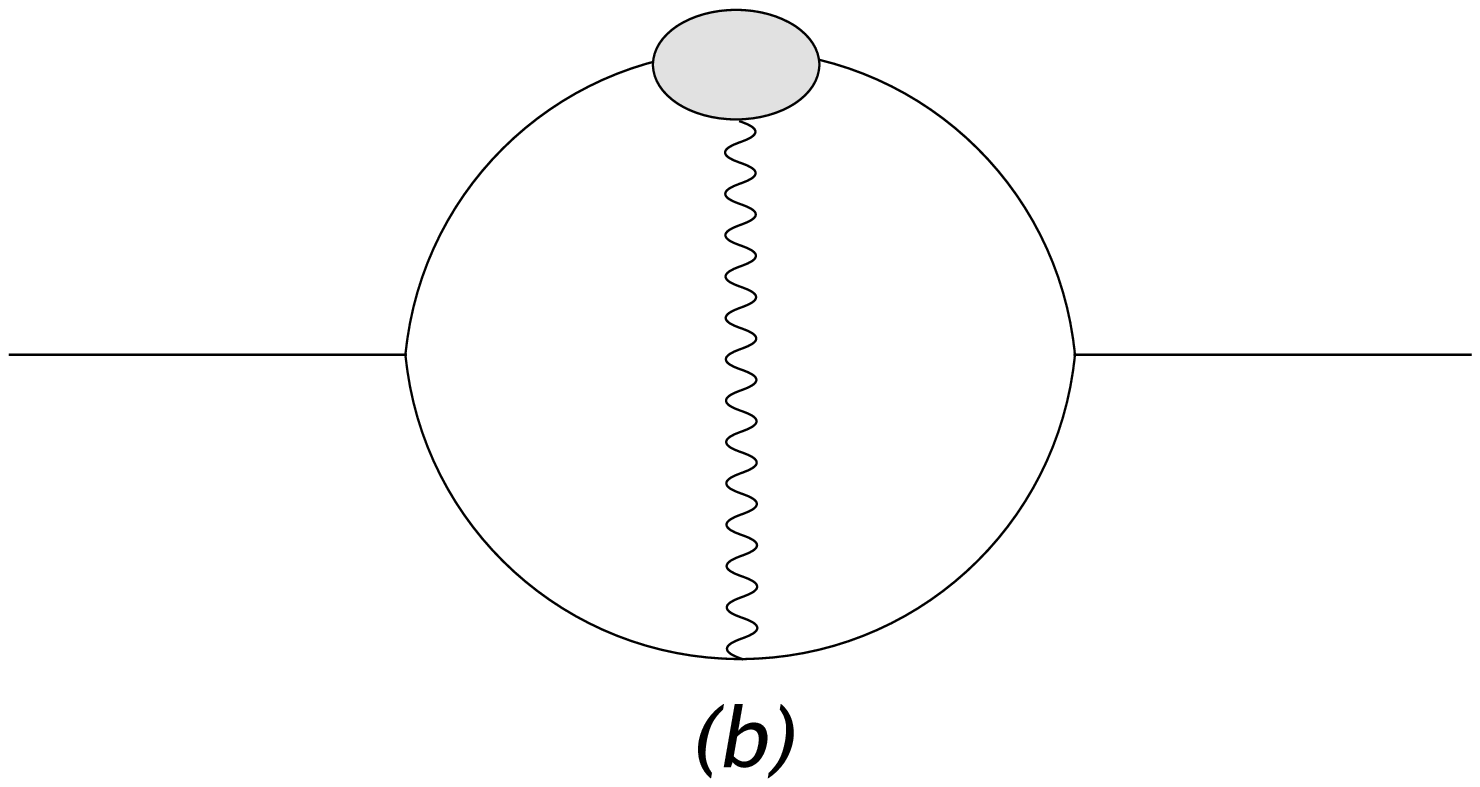} &
\raisebox{-0.07cm}{\epsfysize=2.4cm\epsfbox{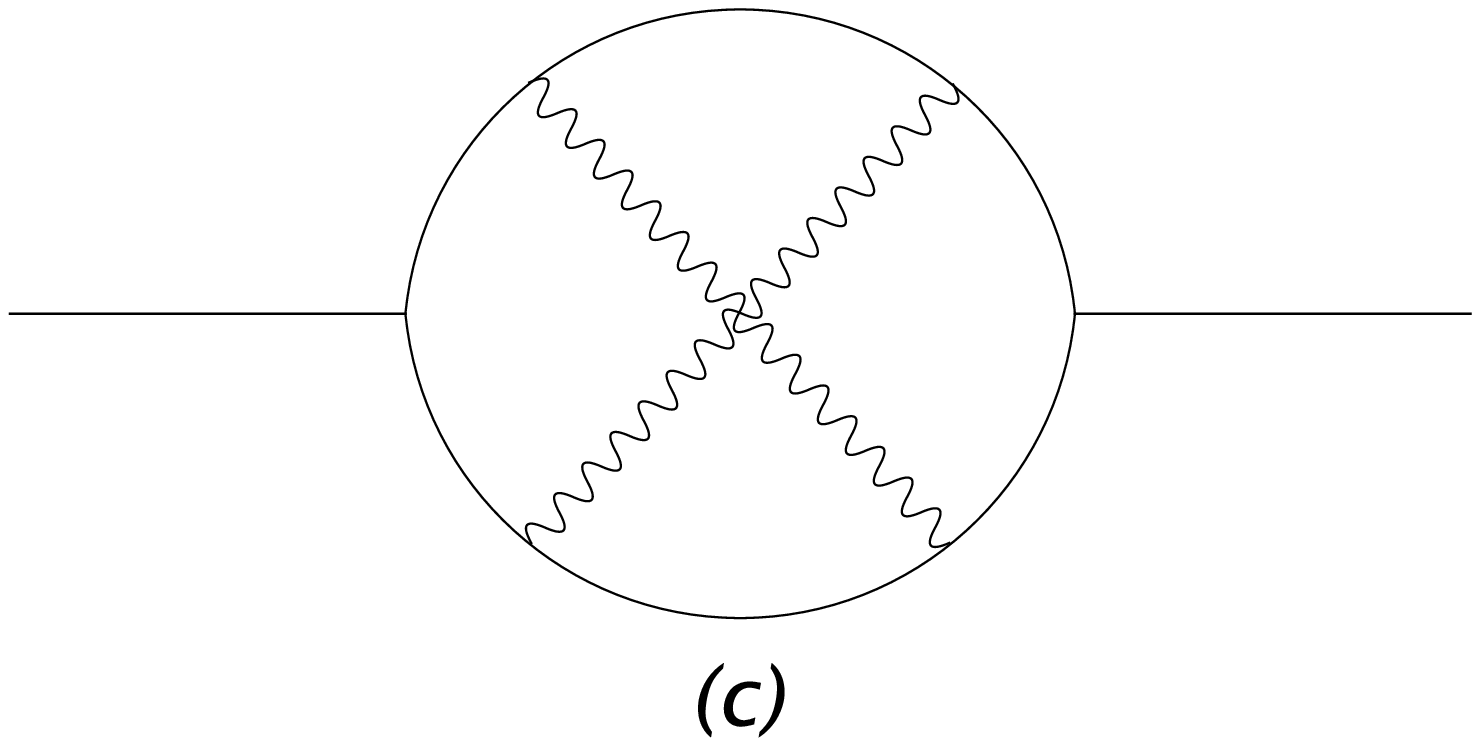}}\\
\epsfysize=2.4cm\epsfbox{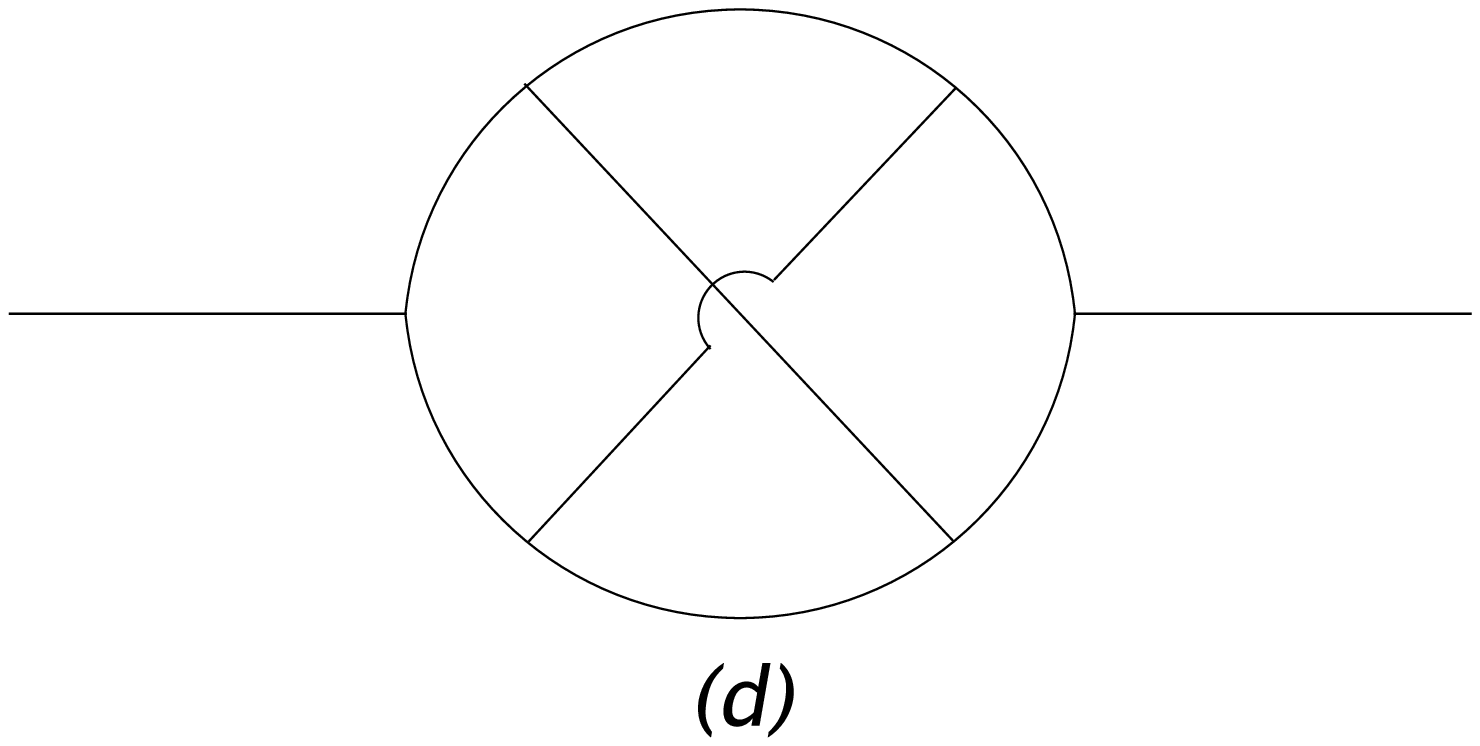} & \raisebox{0.04cm}{\epsfysize=2.4cm\epsfbox{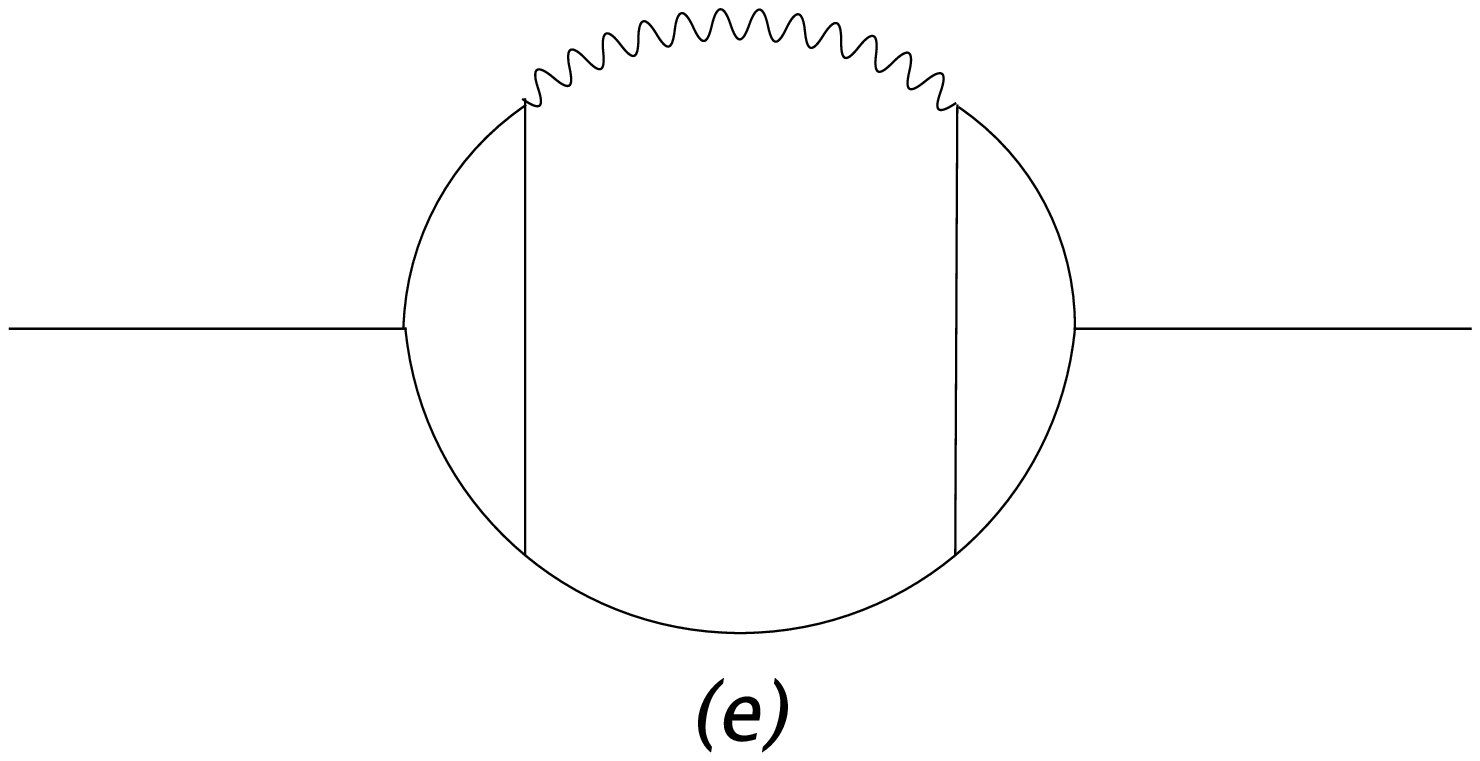}}
\end{tabular}
\end{center}
\caption{Two--loop diagrams for $\langle\mc O_{11} \bar{\mc O}_{11} \rangle$
and $\langle\mc O_{23} \bar{\mc O}_{23} \rangle$.}
\label{2pt2loop}
\end{figure}

At two loops the topologies of diagrams which contribute to $\langle \mc O_{11}
\bar{\mc O}_{11}\rangle$ and $\langle \mc O_{23} \bar{\mc O}_{23}\rangle$
are the ones in Fig. \ref{2pt2loop}.

Here the grey bullets indicate two--loop corrections to the chiral
propagator and one--loop corrections to the mixed gauge--chiral
vertex. Using the superconformal condition (\ref{superconformal2})
their $q,h, h'$ dependence disappears and these corrections
coincide with the ones of the ${\cal N}=4$ theory
\cite{GRS,PSZ1,PSZ2}. Therefore the first three diagrams give the
same kind of contribution to both correlators.

The last two
diagrams contain chiral vertices and they instead differ in the two cases for the number
of $h$ vs. $h'$ insertions: Diagram \ref{2pt2loop}d) gives contributions proportional to
$|h|^4$ and $|h'|^4$ to $\langle\mc O_{11} \bar{\mc O}_{11} \rangle$, and contributions
proportional to $|h|^4$ and $|h|^2 |h'|^2$ to $\langle\mc O_{23} \bar{\mc O}_{23} \rangle$.
Analogously, diagram \ref{2pt2loop}e)
contributes to $\langle\mc O_{11} \bar{\mc O}_{11} \rangle$ with a term proportional to
$g^2 |h'|^2$ and to $\langle\mc O_{23} \bar{\mc O}_{23} \rangle$ with $g^2 |h|^2$.

Diagrams contributing to the mixed two--point function $\langle\mc O_{11} \bar{\mc O}_{23} \rangle$
at two loops are of the type \ref{2pt2loop}d) with two $h$ and two $h'$ vertices
(contributions proportional to  $\bar{h}^2 h'^2$), with three $h$ and one $h'$ (contributions
proportional to $|h|^2 \bar{h}'h$) and \ref{2pt2loop}e)
with one $h$ and one $h'$ vertices (contributions proportional to $g^2 h \bar{h}'$).

Performing the $D$--algebra and computing the corresponding loop integrals
in momentum space and dimensional regularization, it is easy to verify
that the diagrams \ref{2pt2loop}a)--d) have at most $1/\e$ poles which correspond to finite
corrections to the two--point functions when transformed back to the configuration space.

The only potential source of anomalous dimension terms would be the graph \ref{2pt2loop}e) since,
after D--algebra,
the corresponding integral has a $1/\e^2$ pole, that is
a $\log{(\mu^2 x^2)}$ divergence in configuration space. However, when computing the correlators
$\langle\mc O_{11} \bar{\mc O}_{11} \rangle$ and $\langle\mc O_{11} \bar{\mc O}_{23} \rangle$
this diagram gives a vanishing color
factor, whereas for the third correlator there is a complete cancellation between
the contribution corresponding to a particular configuration of the $\bar{\Phi}_2$, $\bar{\Phi}_3$ lines
coming out from the $\bar{\mc O}_{23}$ vertex
and the one with the two lines interchanged (the same happens in the $h'=0$ theory \cite{PSZ4}).

Therefore, all the correlators in configuration space are two--loop finite, the anomalous
dimension matrix vanishes and the two operators are protected up to this order.

It is interesting to give the explicit result for the two--loop corrections to the
correlators. We find
\bea
\langle\mbox{Tr}(\Phi_1^2)(z_1)\mbox{Tr}(\bar{\Phi}_1^2)(z_2)\rangle_{\rm 2-loops}
&\sim& \frac{\delta^{(4)}(\theta_1-\theta_2)}{[(x_1-x_2)^2]^2} \, \mc F_1
\non \\
\langle\mbox{Tr}(\Phi_2\Phi_3)(z_1)\mbox{Tr}(\bar{\Phi}_2\bar{\Phi}_3)(z_2)\rangle_{\rm 2-loops}
&\sim& \frac{\delta^{(4)}(\theta_1-\theta_2)}{[(x_1-x_2)^2]^2} \, \mc F_2
\eea
where
\bea\label{a1}
\mc F_1&=&\left[ |h|^4\frac{N^2-4}{N^2}\left|q-\bar{q}\right|^2\left(\frac{N^2-1}{4N^2}
\left|q-\bar{q}\right|^2-1\right) \right. \non\\
&+&\left. |h'|^4\frac{(N^2-20)(N^2-4)}{8N^4}-|h|^2|h'|^2\frac{N^2-4}{2N^2}
\left(1-\frac{1}{N^2}\left|q-\bar{q}
\right|^2\right)\right]
\eea
and
\bea\label{a2}
\mc F_2&=&\Bigg[|h|^4\frac{N^2-4}{4N^4}\left|q-\bar{q}\right|^4
+|h'|^4\frac{(N^2-4)^2}{8N^4} \non\\
&~&~~~~+~|h|^2|h'|^2\frac{N^2-4}{2N^2}\left(3-\frac{N^2-5}{N^2}\left|q-\bar{q}\right|^2\right)
\Bigg]
\eea
We note that all the $g^4$ contributions cancel and we are left with expressions which vanish
in the ${\cal N}=4$ limit ($\b = h' = 0$, $|h|^2 = g^2$).
Moreover, both the contributions survive in the large $N$ limit in contradistinction to the
$h'=0$ case where $\mc F_2$ is subleading \cite{PSZ4}.

\subsection{Chiral ring: The  $\D_0 = 3$ sector}

The next sector we investigate contains operators with naive scale
dimension $\Delta_0=3$. We classify them according to their $Z_3$--charge as in Table 2.

\begin{table}[ht]
\begin{center}
\begin{tabular}{|*{3}{c|}|}
\hline
$\mc Q=0 $ & $\mc Q=1 $ & $\mc Q=2 $ \\
\hline
\hline
$\mc O_1=\mbox{Tr}(\Phi_1^3)~~~~~~~$ & $\mc O_6=\mbox{Tr}(\Phi_1^2\Phi_2)$ & $\mc O_9=\mbox{Tr}(\Phi_1^2\Phi_3)$\\
\hline
$\mc O_2=\mbox{Tr}(\Phi_2^3)~~~~~~~$ &  $\mc O_7=\mbox{Tr}(\Phi_2^2\Phi_3)$ & $\mc O_{10}=\mbox{Tr}(\Phi_3^2\Phi_2)$ \\
\hline
$\mc O_3=\mbox{Tr}(\Phi_3^3)~~~~~~~$ &  $\mc O_8=\mbox{Tr}(\Phi_3^2\Phi_1)$ & $\mc O_{11}=\mbox{Tr}(\Phi_2^2\Phi_1)$ \\
\hline
$\mc O_4=\mbox{Tr}(\Phi_1\Phi_2\Phi_3)$ & ~~ & ~~ \\
\hline
$\mc O_5=\mbox{Tr}(\Phi_1\Phi_3\Phi_2)$ & ~~ & ~~ \\
\hline
\end{tabular}
\end{center}
\caption{Operators with $\Delta_0=3$.}
\label{opw3}
\end{table}

We note that the neutral sector does not contain the same number of
operators as the charged ones. This is due to the fact that, in contradistinction to
the previous case, the $\mc Q=0$ sector is closed under the
application of cyclic permutations $\Phi_i \rightarrow \Phi_{i+1}$ and
tranformations (\ref{qsymmetry}). Therefore, we cannot generate the
charged sectors from the neutral one by using these mappings.

The charged sectors are also closed under permutations but they
get exchanged under transformations (\ref{qsymmetry}). This is the
reason why they still have the same number of operators.

We first focus on the set of operators with $\mc Q=0$.
As for the $h'= 0$ theory, in this sector the Konishi anomaly enters
the game when we try to use the equations of motion to write descendants
which involve $\mc O_4$ and $\mc O_5$. However, as discussed in Section 6,
the Konishi anomaly can be neglected as long as we are interested in the
construction of CPO's up to two loops. We will then restrict our analysis
at this order.

Using
the equations of motion (\ref{eom2}) we can write three descendant
operators \bea\label{desc4}
\mc D^{(1)}&=&h\left(q\,\mc O_4-\bar{q}\,\mc O_5\right)+h'\,\mc O_1 \non\\
\mc D^{(2)}&=&h\left(q\,\mc O_4-\bar{q}\,\mc O_5\right)+h'\,\mc O_2 \\
\mc D^{(3)}&=&h\left(q\,\mc O_4-\bar{q}\,\mc O_5\right)+h'\,\mc
O_3 \non \eea
According to the discussion of Section \ref{conj} we
expect to single out two protected operators. We consider the most
general linear combination
\beq \mc P=\a \,\mc O_1+\b \,\mc
O_2+ \g \,\mc O_3+ \d \,\mc O_4+ \e \,\mc O_5
\eeq
and require
tree-level orthogonality to the three descendants. These
constraints provide the condition $\a =\b =\g \equiv a$ (as
expected because of the $Z_3$ symmetries of this sector) and the
extra relation
\beq 3 a
\bar{h}'(N^2-4)q+\bar{h}\left[\d \,\left(N^2-2+2q^2\right)-\e \,\left((N^2-2)q^2+2\right)\right]=0
\eeq
which can be used to express $a$ in terms of two arbitrary constants.

Any CPO in this sector has then the following form
\beq \mc
P=a\,(\mc O_1+\mc O_2+\mc O_3)+\d \,\mc O_4+ \e \,\mc O_5
\label{cpo} \eeq
An explicit check on its two--point function at
one loop leads to  $\langle\mc P\,\bar{\mc P}\rangle_1$ finite,
independently of the choice of $\d$ and $\e$. One can choose
the two constants in order to select two mutually orthogonal
operators.

As it happened in the previous cases, these operators are guaranteed to be protected up to two loops as a
consequence of their one--loop protection plus the result $W_{eff}^{(1)} \sim W$ which insures
that the classical
descendants (\ref{desc4}) keep being good descendants also at one loop.

The sectors characterized by $Z_3$ charges $\mc Q=1,2$ do not
contain protected operators. In fact, one can see that any charged operator in
Table 2 can be written as $\mc O_i= \bar{D}^2X_i$ by using the classical equations of motion. We
expect this result to be valid at any order of perturbation theory
since the structure of the effective superpotential for what
concerns its superfield dependence cannot change.

To summarize, in the $\D_0 = 3$ sector we have found two protected operators which are linear
combinations
of ${\rm Tr}(\Phi_i^3)$, $i=1,2,3$, ${\rm Tr}(\Phi_1\Phi_2\Phi_3)$ and ${\rm Tr}(\Phi_1\Phi_3\Phi_2)$.
We note that
among all possible weight--3 operators these are the only ones which belong to the chiral ring of the
$\beta$--deformed
theory. The rest of weight--3 operators which were descendants for $h'=0$ keep being descendants.

The protected operators we have found are neutral under the $Z_3$
symmetry (\ref{Z3}). As discussed in \cite{BJL}, the neutral
sector of the chiral ring (the untwisted sector) coincides with
the center of the quantum algebra generated by the $F$--terms
constraints. In particular, for the $h'$--deformation one element
of the center has been constructed explicitly (eq. (4.83) in
\cite{BJL}). This element coincides with one of the CPO's
(\ref{cpo}) we have found, once we set $\mc D^{(i)} =0$ in the
chiral ring (see eq. (\ref{desc4})), use these identities to
express the operator $\mc O_5$ in terms of the other ones and make
a suitable choice for the coefficients $\d$ and $\e$.

\subsection{Comments on the general structure of the chiral ring}

The $\D_0 = 2,3$ sectors studied in the previous Section are very
peculiar and do not provide enough informations to guess the
structure of the sectors for generic scale dimension. In fact, for
$\D_0 =2$ no descendants are present and we cannot even apply the
orthogonality procedure to construct CPO's. The $\D_0=3$ sector
contains only protected operators which are $Z_3$ neutral and are
linear combinations of ``old'' CPO's, that is operators which were
protected for $h'=0$.

A naive generalization of our results to higher dimensional
sectors would lead to the conjecture that the chiral ring for the
$h'$--deformed theory, at least for what concerns its neutral
sector with $\D_0 =3J$, would be given by linear combinations of
${\rm Tr}(\Phi_i^{3J})$ and ${\rm Tr}(\Phi_1^J\Phi_2^J\Phi_3^J)$.
However, we expect more general operators of the form ${\rm
Tr}(\Phi_1^{3J-m-n} \Phi_2^m \Phi_3^n)$, $m+2n= {\rm mod}(3)$ to
appear. Moreover, nontrivial $Z_3$--charged sectors should
appear for $\D_0 = 3J$ even if they are absent in the particular
case $\D_0 =3$.

To investigate these issues we should extend our analysis to
higher dimensional sectors and this would require quite a bit of
technical effort. However, without entering any calculative
detail, but simply performing dimensional and $Z_3$--charge
balances we can figure out few general properties of the $\mc
Q$--sectors of the chiral ring.

We consider the generic chiral operator $\mc O_1 = (\Phi_1^a \Phi_2^b
\Phi_3^c)$ for any trace structure with scale dimension $\D_0=
a+b+c$ and $Z_3$--charge $\mc Q_1 \equiv b + 2c $ with respect to the
symmetry (\ref{Z3}).

We now perform $\Phi_i \leftrightarrow \Phi_j$ exchanges according
to the symmetry (\ref{qsymmetry}) and $Z_3$ permutations. In this
way of doing we generate all the operators with the same trace
structure in a given $\D_0$ sector. Let us consider for example
the operators $\mc O_2 = (\Phi_2^a \Phi_1^b \Phi_3^c)$ and $\mc
O_3 = (\Phi_3^a \Phi_1^b \Phi_2^c)$ obtained by a $\Phi_1
\leftrightarrow \Phi_2$ exchange and a cyclic permutation,
respectively. They have charges $\mc Q_2 = a+2c$ and $\mc Q_3 = 2a
+c$. It is easy to see that if $\D_0 = 3J$ then $\mc Q_2 = \mc Q_3
= 0 ~ ({\rm mod}(3))$ {\em iff}  $\mc Q_1 = 0 ~ ({\rm mod}(3))$.
This property holds for any operator that we can construct from
$\mc O_1$ by the application of the two discrete symmetries. On
the other hand, if $\mc Q_1 = 1,2 ({\rm mod}(3))$ operators
obtained from it by cyclic permutations still mantain the same
charge, but the application of field exchanges (\ref{qsymmetry})
map charge--1 operators into charge--2 operators and viceversa.

Therefore, for $\D_0=3J$ the $\mc Q=0$ class is closed under the
action of $Z_3$--permutations and (\ref{qsymmetry}) symmetry, and
being independent, may contain a different number of operators
compared to the charged sectors which instead are related by
(\ref{qsymmetry}) mappings. In particular, as it happens for
$\D_0=3$ charged classes of the chiral ring might be empty while
the corresponding neutral one is not.

If $\D_0 \neq 3J$ a simple calculation leads to the
conclusion that starting from operators with zero $Z_3$--charge we
generate operators with $\mc Q = 1$ by applying $\Phi_1
\leftrightarrow \Phi_2$ if $\D_0 = 3J+1$ and a cyclic permutation
if $\D_0 = 3J+2$. Correspondingly, we obtain operators with $\mc Q = 2$
by applying a cyclic permutation in the first case and
a $\Phi_1 \leftrightarrow \Phi_2$ exchange in the second case. Therefore, in
any sector with $\D_0 \neq 3J$ the number of operators with $\mc Q
= 1$ is the same as the ones with $\mc Q = 2$ and coincides with
the number of neutral operators.

If we apply the same reasoning to the descendant operators of each
sector (to simplify the analysis we work at large $N$ to avoid
mixing among different trace structures) we discover that every time
$\D_0 \neq 3J$ the descendants of the charged classes can be
obtained from the neutral ones by field exchanges. As a
consequence, the three classes contain the same number of
descendants and then the {\em same} number of
protected operators.

To summarize, the sectors of the chiral ring behave differently
according to their scale dimension: If $\D_0 \neq 3J$ the
corresponding operators are equally split into the three $\mc Q$
classes. On the contrary, if $\D_0 = 3J$ the neutral class is
independent and may contain a different number of CPO's.

As a further example we have studied the $\D_0=4$ operators. In
the large $N$ limit and at the lowest order in perturbation theory
we have found that the neutral single--trace sector contains one independent CPO
(we have eight single--trace chirals and seven descendants).
Therefore, we conclude that also the charged sectors contain one
single protected operator and we know how to construct it once we
have found the $\mc Q=0$ operator explicitly. In the single--trace
sector the protected operator turns out to be a linear combination of
\bea
&& {\rm Tr}(\Phi_1^4)
\non \\
&& {\rm Tr}(\Phi_1 \Phi_2^3) \quad , \quad {\rm Tr}(\Phi_1
\Phi_3^3) \quad , \quad  {\rm Tr}(\Phi_2^2 \Phi_3^2) \quad , \quad
{\rm Tr}(\Phi_2 \Phi_3 \Phi_2 \Phi_3)
\non \\
&& {\rm Tr}(\Phi_1^2 \Phi_2 \Phi_3) \quad , \quad {\rm
Tr}(\Phi_1^2 \Phi_3 \Phi_2) \quad , \quad {\rm Tr}(\Phi_1 \Phi_2
\Phi_1 \Phi_3)
\eea

It remains the open question whether for $\D_0 = 3J$, $J >1$, the
charged sectors are trivial as in the weight--3 case. A systematic
analysis of the charged protected operators is a difficult task in
general. However, working at large $N$ it is easy to realize that
for $J$ {\em even} and $J > 1$, there are nontrivial protected operators
for $\mc Q=1$ and $\mc Q =2$. These are operators with the
$3J$ chiral superfields split into the maximal number of traces
allowed by $SU(N)$, i.e. $3J/2$. In fact, for these operators it
is impossible to exploit the equations of motion and write them as
descendants. For $J$ {\em odd} the same arguments do not lead to any
definite conclusion. However, we expect to generate nontrivial
charged protected operators by multiplying the neutral CPO's of
weight 3 previously constructed by $3(J-1)/2$ traces containing two
operators each and carrying the right $Z_3$ charge.

\section{Conclusions}

In this paper we have considered ${\cal N}=1$ $SU(N)$ SYM theories obtained as
marginal deformations of the ${\cal N}=4$ theory.
In particular, we have focused on the perturbative structure of the matter (not gauge)
quantum chiral ring defined as in (\ref{quantumring}) in terms of the effective
superpotential. According to our general prescription, CPO's can be determined by imposing order
by order the orthogonality condition (\ref{cond}) to all the descendants of a given
sector. This requires constructing first the descendants as a power expansion in the
couplings. According to the definition (\ref{quantumring}), this can be easily accomplished
once the effective superpotential is known at a given order.

For the Lunin--Maldacena $\b$--deformed theory (\ref{baction}) we
have studied quite extensively the spin--2 sector of the theory.
For the particular examples of weights $(J,1,0)$ and $(2,2,0)$ we
have considered, a special pattern arises which allows for a
drastic simplification in the study of the orthogonality
condition: In any of these sectors descendants can be always
constructed at tree level which turn out to be good independent
descendants even at the quantum level. This is due to the
particular form (\ref{Weff}) of the superpotential and the
peculiar way the equations of motion work which allow for
constructing $q$--independent descendants, insensible to the
quantum corrections of the theory. This property persists even for
other examples of the form $(J_1,J_2,0)$. Therefore, we conjecture
that it might be a property of the entire spin--2 sector: For
any weight $(J_1,J_2,0)$ quantum descendant operators can be
constructed which coincide with the descendants determined
classically.

We have then studied the spin--3 sector. In this case the
determination of quantum descendants of weights $(J_1,J_2,J_3)$
cannot ignore the Konishi anomaly term. Being its effect of order
$\l$ it only enters nontrivially the orthogonality condition from
two loops on, that is it will affect the form of the protected
operators at least at three loops. For weights $(1,1,1)$ and
$(2,1,1)$ we have determined the CPO's up to two loops. In
particular, for the first case we have proved that up to this
order the correct CPO is the one found in \cite{FG}. Higher order
calculations would require computing two--point correlation
functions between matter chiral operators and ${\rm Tr}(W^\a
W_\a)$. It would be interesting to pursue this direction since it
represents the first case where the descendant operators, apart
from acquiring an explicit dependence on the Konishi anomaly term,
get modified nontrivially at the quantum level due to the nontrivial
corrections to the superpotential which start appearing at order
$\l^2$.

We have extended our procedure to the study of protected operators for the full
Leigh--Strassler deformation. We can think of this theory as a marginal
perturbation of the $\b$--deformed theory induced by the $h'$--terms in (\ref{action}).
In this case the determination of the
complete chiral ring is a difficult task and only few insights have been discussed in
\cite{BJL}. We have moved few steps in this direction by studying perturbatively
the simple $\D_0 = 2,3$ sectors. For operators of scale dimension two we have found
that the $h'$--deformed theory has still the same CPO's as the $h'=0$ one,
i.e. ${\rm Tr}(\Phi_i^2)$ and ${\rm Tr}(\Phi_i \Phi_j)$, $i \neq j$.

For the $\D_0 =3$ sector we have found a two--dimensional plane of
CPO's given as linear combinations of the CPO's of the
corresponding $h'=0$ theory, i.e. ${\rm Tr}(\Phi_i^{3})$ and ${\rm
Tr}(\Phi_1\Phi_2\Phi_3)$. In fact, in this case the lower number
of global symmetries surviving the deformation allows for mixing
among the operators who were protected in the previous case and
belonged to different $U(1) \times U(1)$ sectors. The class of
protected operators we have found contains the central element of
the quantum algebra proposed in \cite{BJL}.

What turns out is that in the $\D_0 =2$ sector the chiral ring is
made by operators which are both charged and neutral with respect to
the $Z_3$--symmetry (\ref{Z3}) that the theory inherits from the
parent $h'=0$ theory. On the other hand, in the $\D_0 =3$ sector
{\em all} CPO's we can construct are neutral under (\ref{Z3}). The
generalization of our results to higher dimensional sectors leads
to the result that the chiral ring for the $h'$--deformed theory
can be divided into two subsets: Sectors with scale dimension
$\D_0 = 3J$ have an independent $\mc Q =0$ class which may contain
in general a different number of CPO's. Instead, whenever $\D_0
\neq 3J$ we can generate the chiral primary operators of the
charged classes from neutral CPO's by the use of the other
discrete symmetries, i.e. cyclic permutations of the three
superfields and the symmetry (\ref{qsymmetry}). It then follows
that the three classes contain the same number of protected
operators. In particular, for any non--empty neutral sector (for
instance $\D_0=2,4$) the corresponding charged ones are
nontrivial. Neutral CPO's will be in general linear combinations
of operators of the form ${\rm
Tr}(\Phi_1^{J-m-n}\Phi_2^m\Phi_3^n)$ with $m+2n= 3p$.

The $Z_3$ periodicity we have found in the chiral ring structure
should have a counterpart in the spectrum of BPS states of the
dual supergravity theory. Therefore, it might be of some help
in the construction of the dual spectrum.

For {\em all} the cases we have investigated the CPO's do not get
corrected at one--loop, whereas they start being modified at order
$\l^2$. This one--loop non--renormalization found for a large
class of chiral operators is probably universal for all the CPO's
and might be traced back to the one--loop non--renormalization
properties of the theories. Precisely, the conditions
(\ref{superconformal}, \ref{superconformal2}) which insure
superconformal invariance at one--loop are maintained at two
loops, i.e. the superconformal theories at one and two loops are
the same. It is then natural to speculate that the corresponding chiral rings
should be the same. The theory instead changes at three loops where the
superconformal condition gets modified by terms of order $\l^2$
\cite{RSS}. Therefore we expect that at this order
the chiral ring will be modified by effects of the same order.

\vspace{1.5cm}

\section*{Acknowledgements}

\noindent This work has been supported in
part by INFN, PRIN prot. $2005-024045-004$ and the European Commission RTN
program MRTN--CT--2004--005104.

\newpage
%%%%%%%%%%%%--Appendices--%%%%%%%%%%%%%%%%%%%%%

\appendix

\section{Integrals in momentum space}\label{inte}

In this Appendix we list the results for loop integrals that we have
used along the calculations.
Working in momentum space and dimensional regularization ($n=4-2\epsilon$)
we give the results as $\epsilon$ expansions.

We begin by considering the momentum integrals associated to the
one--loop and two--loop diagrams in Fig. \ref{effsup} for the
perturbative corrections to the superpotential.

At one loop, after performing $D$--algebra, the diagram \ref{effsup}b)
gives the standard triangle contribution \cite{UD}. Assigning external momenta
$p_i$ ($p_1+p_2+p_3=0$) we have
\bea
p_3^2 \int\frac{d^nq}{(2\pi)^n}\frac{1}{q^2(q-p_2)^2(q+p_1)^2}=
\frac{1}{(4\pi)^2} \, \Phi^{(1)}\left(x,y\right)+\mc O(\epsilon)
\eea
where
\beq
\label{xy}
x\equiv\frac{p_1^2}{p_3^2}\,\,\,\,\,\,\,\mbox{and}\,\,\,\,\,\,\,y\equiv\frac{p_2^2}{p_3^2}
\eeq
The $p_3^2$ in front of the integral is produced by $D$--algebra.
The function $\Phi^{(1)}(x,y)$ can be represented as a parametric integral
\beq
\Phi^{(1)}(x,y)=-\int^1_0\frac{d\xi}{y\,\xi^2+(1-x-y)\xi+x}\left(\log{\frac{y}{x}}+2\log{\xi}\right)
\eeq
Since we look for a local contribution to the superpotential we are interested in
the result of the integral for external momenta set to zero.
A consistent way \cite{west} to take the limit of vanishing external momenta
is to set $p_i^2=m^2$ for any $i$
so having $x,y = 1$ and let the IR cut--off $m^2$ going to zero at the end of the calculation.
In the limit we obtain a finite local result \cite{west}
\beq
- \int_0^1 d\xi \frac{\log{\xi (1-\xi)}}{1 - \xi(1-\xi)}
\eeq
At two loops two types of integrals appear. From diagrams \ref{effsup}c) and \ref{effsup}d)
we have integrals of the form
\bea
&&(p_3^2)^2 \int\frac{d^nq\,d^nr}{(2\pi)^{2n}}
\frac{1}{(r+p_1)^2(q+p_1)^2(r-p_2)^2(q-p_2)^2r^2(q-r)^2}=\non\\
&=&\frac{1}{(4\pi)^4} \, \Phi^{(2)}\left(x,y\right)+\mc O(\epsilon)
\eea
with $x$ and $y$ as in (\ref{xy}). The function $\Phi^{(2)}(x,y)$ is defined by \cite{UD}
\beq
\Phi^{(2)}(x,y)=-\frac{1}{2}\int^1_0\frac{d\xi}{y\,\xi^2+(1-x-y)\xi+x}\log{\xi}\left(\log{\frac{y}{x}}+
\log{\xi}\right)\left(\log{\frac{y}{x}}+2\log{\xi}\right)
\eeq
As in the one--loop case, the limit $x,y\rightarrow 1$ gives a finite local contribution to
the effective superpotential.

From diagrams \ref{effsup}c)--g) this kind of integral also
appears
\bea
p_3^2\int\frac{d^nq\,d^nr}{(2\pi)^{2n}}\frac{1}{q^2\,r^2(q-r)^2(q-p_3)^2(r-p_3)^2}=
\frac{1}{(4\pi)^4}\,6\zeta(3) +\mc O(\epsilon)
\eea
where one of
the external momenta has been already set to zero (in this case we
can safely set one of the external momenta to zero from the very
beginning since we do not introduce fake IR divergences). This is
already the local finite contribution we obtain by setting also
$p_3^2 =0$.

\vskip 15pt
When we deal with two-point correlation functions, at tree-level we have ($k=\Delta_0$
is the free scale dimension of the operators involved and $p$ is the external momentum)
\bea
&~&\int\frac{d^nq_1\,...\,d^nq_{k-1}}{(2\pi)^{n(k-1)}}\frac{1}{q_1^2(q_2-q_1)^2(q_3-q_2)^2...
(p-q_{k-1})^2} \non\\
&~&~~~~~~~~~~~ = ~\frac{1}{\epsilon}\left[\frac{1}{(4\pi)^2}\right]^{k-1}
\frac{(-1)^k}{[(k-1)!]^2}(p^2)^{k-2-(k-1)\epsilon}+\mc O(1)
\eea

At two loops we are interested in the four diagrams listed in
Fig. \ref{cross}. From the graph \ref{cross}a) we obtain
\bea
&~&\int\frac{d^nq_3\,...\,d^nq_{k-1}}{(2\pi)^{n(k+1)}}\frac{1}{(q_4-q_3)^2...
(p-q_{k-1})^2} \times \non\\
&~&~~~~\int\frac{d^nk\,d^nl\,d^nr\,d^ns\,\,\,\,\,\,\,
r^2(q_3-l)^2}{k^2l^2(k-l)^2(r-k)^2(r-l)^2(s-l)^2(r-s)^2(q_3-r)^2(q_3-s)^2} \\
&& ~~~\non \\
&~&~~~~~~~~~= ~\frac{1}{\epsilon}\left[\frac{1}{(4\pi)^2}\right]^{k+1}
\frac{(-1)^k(k-1)}{[(k-1)!]^2(k+1)}[6\zeta(3)-20\zeta(5)](p^2)^{k-2-(k+1)\epsilon}+\mc O(1) \non
\eea
The momentum integral for the graph \ref{cross}b) gives
\bea
&~&\int\frac{d^nq_3\,...\,d^nq_{k-1}}{(2\pi)^{n(k+1)}}
\frac{-q_3^2}{(q_4-q_3)^2...(p-q_{k-1})^2} \times \non\\
&~&~~~~\int\frac{d^nk\,d^nl\,d^nr\,d^ns}{k^2l^2(k-l)^2(r-k)^2(s-l)^2(r-s)^2(q_3-r)^2(q_3-s)^2}
\\
&& ~~~~~\non\\
&~&~~~~~~~~ =~\frac{1}{\epsilon}\left[\frac{1}{(4\pi)^2}\right]^{k+1}
\frac{(-1)^k(k-1)}{[(k-1)!]^2(k+1)}40\zeta(5)(p^2)^{k-2-(k+1)\epsilon}+\mc O(1) \non
\eea
Finally, the graphs \ref{cross}c) and \ref{cross}d) lead to the same contribution
\bea
&~&\int\frac{d^nr\,d^nq_2\,...\,d^nq_{k-1}}{(2\pi)^{n(k+1)}}
\frac{1}{(q_2-r)^2(q_3-q_2)^2...(p-q_{k-1})^2} \times \non\\
&~&~~~~\int\frac{d^nk\,d^nl}{k^2l^2(k-l)^2(r-k)^2(r-l)^2} \\
&& ~~~~~\non\\
&~&~~~~~~~~ =~ \frac{1}{\epsilon}\left[\frac{1}{(4\pi)^2}\right]^{k+1}
\frac{(-1)^k(k-1)}{[(k-1)!]^2(k+1)}6\zeta(3)(p^2)^{k-2-(k+1)\epsilon}+\mc O(1) \non
\eea

\newpage
%%%%%%%%%%%%--References--%%%%%%%%%%%%%%%%%%%%%

\end{document}